\documentclass[conference,10pt]{IEEEtran}
\usepackage{algorithm}
\usepackage{algorithmic}

\usepackage{cite}
\usepackage{epsfig}
\usepackage{epstopdf}
\usepackage{graphicx}
\usepackage{xcolor}
\usepackage{tikz}
\usetikzlibrary{patterns}
\usetikzlibrary{arrows,positioning,shapes.geometric,circuits.logic.US,spy}
\usepackage{pgfplots}
\usepackage{multirow}
\usepackage{upgreek}
\usepackage{amssymb}
\usepackage{amsmath}
\usepackage{cases}
\usepackage{url}
\usepackage{pifont}
\usepackage{bm}
\usepackage{amsthm}

\newtheorem{proposition}{Proposition}
\newtheorem{lemma}{Lemma}

\usepackage{tabularx}
\usepackage{booktabs}
\usepackage{filecontents}
\usepackage{subcaption}
\newcolumntype{Y}{>{\centering\arraybackslash}X}

\usepackage{array}

\newcommand{\fixme}[2]{\ifx&#2&{\leavevmode\color{red}#1}\else{\leavevmode\color{red}FIXME\{}#1{\leavevmode\color{red}\}}\footnote{{\leavevmode\color{red}#2}}\PackageWarning{Fixme}{#1: #2}\fi}

\newcommand{\newstuff}[2]{\ifx&#2&{\leavevmode\color{blue}#1}\else{\leavevmode\color{blue}FIXME\{}#1{\leavevmode\color{blue}\}}\footnote{{\leavevmode\color{blue}#2}}\PackageWarning{Newstuff}{#1: #2}\fi}

\DeclareMathOperator*{\sgn}{sgn}
\DeclareMathOperator{\PM}{PM}

\DeclareMathOperator*{\row}{row}
\DeclareMathOperator*{\vect}{vec}

\title{Practical Product Code Construction of Polar Codes}

\author{\IEEEauthorblockN{Carlo~Condo, Valerio Bioglio, Hartmut Hafermann, Ingmar~Land\\}
\IEEEauthorblockA{Mathematical and Algorithmic Sciences Lab\\ Huawei Technologies France SASU \\
Email: $\{$carlo.condo,valerio.bioglio,hartmut.hafermann,ingmar.land$\}$@huawei.com}} 
\begin{document}

\maketitle
\begin{abstract}
In this paper, we study the connection between polar codes and product codes. 
Our analysis shows that the product of two polar codes is again a polar code, and we provide guidelines to compute its frozen set on the basis of the frozen sets of the component polar codes. 
Moreover, we show how polar codes can be described as irregular product codes. 
We propose a two-step decoder for long polar codes taking advantage of this dual nature to heavily reduce decoding latency. 
Finally, we show that the proposed decoding technique outperforms both standard polar codes and state-of-the-art codes for optical communications under latency constraints. 
\end{abstract}

%\begin{IEEEkeywords}
%
%\end{IEEEkeywords}

\IEEEpeerreviewmaketitle

\section{Introduction}\label{sec:intro}

Polar codes \cite{arikan} are capacity-achieving linear block codes that rely on channel polarization.
This phenomenon creates virtual single-bit channels that are either completely noisy or completely noiseless under successive cancellation (SC) decoding as the code length tends to infinity. 
Given its poor error-correction performance for polar codes of moderate code lengths,  list decoding was proposed in \cite{tal_list} to improve SC performance for practical code lengths; the resulting SC-List (SCL) algorithm exhibits enhanced error-correction performance, at the cost of higher decoding latency and complexity. 
However, the fundamental drawback of SC-based decoding algorithms, namely that they are inherently sequential, binds good error-correction performance to long decoding latency.

On the other hand, product codes \cite{elias} are parallel concatenated codes whose decoding process can be easily parallelized.
This code construction is often used in optical communication systems thanks to its good error-correction performance and high throughput.
To reduce their decoding latency, systematic polar codes have been concatenated with short block codes as well as LDPC codes \cite{pc_inner,BP_pc}, achieving good error correction performance with low latency.
However, the use of two different component codes increases the implementation cost due to the large number of decoders to be instantiated to fully exploit the decoding parallelism. 
To solve this problem, authors in \cite{par_conc_sys_pol} recently propose to use two systematic polar codes in the concatenation scheme, simplifying the decoder structure. 

In this paper, we move a step further by studying the effect of the concatenation of non-systematic polar codes in the construction of product codes. 
This analysis is an extension of our preliminary work presented in \cite{PRODPOL_FIRST}, where it is shown that product codes constructed with non-systematic polar codes can be decoded as a unique polar code. 
In this work, we develop the theory behind the dual interpretation of polar codes as product codes, and detail how to switch between the frozen set of the product code to the frozen set of the polar code, and vice versa. 
We propose a low-complexity soft decision decoder taking advantage of the proposed product polar interpretation to improve the performance of the hard decision decoder presented in \cite{PRODPOL_FIRST} while keeping a low decoding latency. 
We then propose a frozen set selection approach that allows to tune the error-correction performance and latency of the mixed product-polar decoding approach. 
In particular, this selection allows to trade-off the effectiveness of the faster, less powerful product decoding used as a first step and that of the more powerful, slower polar decoding used as post processing. 
Finally, we extend the mixed decoding approach of \cite{PRODPOL_FIRST} to various component code decoding algorithms and information exchange criteria between decoding phases.

%A mixed product-polar decoding approach was introduced, showing that the fast and parallel product decoding can be exploited to decode very long polar code, using the standard polar code decoding as a post-processing to correct residual errors. 
\subsection{Related Works} \label{sec:previous}

The description of polar codes as product polar codes has been implicitly suggested in literature, usually to improve SC decoding and without leveraging on the two-dimensional concatenation to propose an alternative decoding algorithm. 
In \cite{Arikan_2D} Ar{\i}kan proposes to run independent row SC decoders to improve the performance of the full code. 
When these decoders encounter an information bit, computation is stopped; an ML decoder is then used on the column code to set final hard decisions, so that row decoders can continue the decoding.
In practice, Ar{\i}kan is proposing to decode the full polar code using an SC decoder in which some operations are performed with ML decoding. 
This decoding strategy is clearly equivalent to following the SC decoding tree of the code for a certain number of stages and then perform the simultaneous decoding of the children nodes input bits through ML decoding. 
Similarly, authors in \cite{Product2014} study SC decoding by separating it in two smaller decoders.  
Again, row decoders have to stop the decoding at every bit to wait for the feedback of column decoders, making the proposed decoding strategy equivalent to SC decoding of the full code. 
Trifonov in \cite{Trifonov_1} demonstrated that polar codes are a class of generalized concatenated codes, and that successive cancellation decoding is an instance of multistage decoding.
However, the idea of describing polar codes as irregular product codes is not contemplated in the paper. 
In fact, the outer codes given in \cite{Trifonov_1} are not the component codes of the irregular product code, since the frozen sets of columns codes are not provided. 
The author further studies concatenation of polar codes with other channel codes in \cite{Trifonov_3}. 
%Finally, the concatenation of two systematic polar codes has been proposed. 

Authors in \cite{par_conc_sys_pol} propose to use two systematic polar codes in the concatenation scheme in order to simplify the decoder structure. 
Soft cancellation (SCAN) \cite{SCAN_pc} and belief propagation (BP) \cite{BP_pc} can be used as soft-input / soft-output decoders for systematic polar codes, at the cost of increased decoding complexity compared to SC.
Recently, SCL decoding has been proposed as a valid alternative to SCAN and BP \cite{par_sys_list}, while authors in \cite{KoikeAkinoIrregularPT} propose to use irregular systematic polar codes to further increase the decoding throughput. 

%\cite{PRODPOL_FIRST}, where is shown that product codes constructed with non-systematic polar codes can be decoded as a unique polar codes. 
%A low complexity, low-latency mixed product-polar decoding algorithm was proposed, showing that the fast and parallel product decoding can be exploited to decode very long polar codes, using the standard polar code decoding as a post-processing to correct residual errors. 

\section{Preliminaries} \label{sec:prel}

\subsection{Polar Codes} \label{subsec:PC}

A polar code of length $N=2^n$ and dimension $K$ is a linear block code built around the polarization effect of the kernel matrix $T_2 = \left[\begin{smallmatrix} 1&0\\1&1 \end{smallmatrix}\right]$.
The transformation matrix $T_N = T_2^{\otimes n}$, defined as the $n$-fold Kronecker power of the polarization kernel, and the frozen set $\mathcal{F} \subset \{1,\dots,N\}$, with $|\mathcal{F}| = N-K$, are the key ingredients for the construction of the code. 
Encoding is performed as 
\begin{equation}
x = u \cdot T_N \text{,} \label{eq:polarGen}
\end{equation}
where the codeword $x = [x_0,x_1,\ldots,x_{N-1}]$ is calculated on the basis of an input vector $u = [u_0,u_1,\ldots,u_{N-1}]$ having the $N-K$ bits in the positions listed in $\mathcal{F}$ set to zero and the remaining $K$ bits storing the information. 
According to the polarization effect, the frozen set collects the $N-K$ less reliable positions under SC decoding, leaving more reliable entries of the input vector to form the information set $\mathcal{I} = \mathcal{F}^C$. 
Reliabilities are usually calculated via Monte Carlo simulation, by tracking the Batthacharyya parameter, or by density evolution under a Gaussian approximation \cite{polar_const}. 
%The generator matrix $G$ of a polar code is calculated from the transformation matrix $T_N$ by deleting the rows of the indices listed in the frozen set. 

SC decoding has been proposed in \cite{arikan} as a soft-input / hard-output decoder for polar codes. 
This algorithm can be described as a depth-first binary tree search, where priority is given to the left branches. 
Soft decisions flow from the root to the leaves, where bits are estimated and hard decisions are propagated towards the root to improve the estimation quality of the next bits. 
To improve the performance of this algorithm for short codes, the SCL decoder has been proposed in \cite{tal_list}, which maintains $L$ parallel codeword candidates. 
The selection of the correct codeword among the candidates can be performed with the aid of cyclic redundancy check (CRC) concatenated to the the polar code. 
Soft-input / hard-output decoders as BP \cite{BP_pc} and SCAN \cite{SCAN_pc} have been proposed for polar codes, however exhibiting poor tradeoffs between increase in complexity and performance improvement. 

\subsection{Product Codes} \label{subsec:prod}

Product codes were proposed in \cite{elias} to provide a simple and efficient way to construct very long codes on the basis of two or more short component codes. 
Given two systematic\footnote{Component codes are usually systematic in order to simplify the encoding, even if this is not a necessary condition.} linear block codes $\mathcal{C}_r$ and $\mathcal{C}_c$ with parameters $(N_r,K_r)$ and $(N_c,K_c)$ respectively, this technique permits to construct a code $\mathcal{P} = \mathcal{C}_c \times \mathcal{C}_r$ of length $N = N_r N_c$ and dimension $K = K_r K_c$. 
Encoding is performed starting from a $K_c \times K_r$ matrix $U$, containing the $K$ information bits. 
Rows are initially encoded independently using code $\mathcal{C}_r$, then the columns of the resulting $K_c \times N_r$ matrix $U_r$ are encoded using code $\mathcal{C}_c$. 
The result is a $N_c \times N_r$ codeword matrix $X$, where rows are codewords of code $\mathcal{C}_r$ and columns are codewords of code $\mathcal{C}_c$. 
It is worth noting that reversing the encoding order does not change the resulting codeword matrix. 
This encoding procedure can be mathematically described as 
\begin{equation}
\label{eq:prod_enc}
X = G_c^T \cdot U \cdot G_r,
\end{equation}
where $G_r$ and $G_c$ are the generator matrices of codes $\mathcal{C}_r$ and $\mathcal{C}_c$ respectively. 
Generator matrix of $\mathcal{P}$ can be calculated through the Kronecker product of the generator matrices of the two component codes as $G = G_c \otimes G_r$ \cite{mcw_sloane}. 

Decoding is performed by sequentially decoding rows and column component codes while exchanging information between the two decoders. 
Row (column) component codes decoding can be performed concurrently since no information is directly exchanged among rows (columns). 
Soft-input/soft-output algorithms can improve the decoding performance by exchanging soft information \cite{block_turbo}. 

\section{From Product to Polar Codes} \label{sec:bridge}
\begin{figure}[t]
  \centering
  \includegraphics[width=0.45\textwidth]{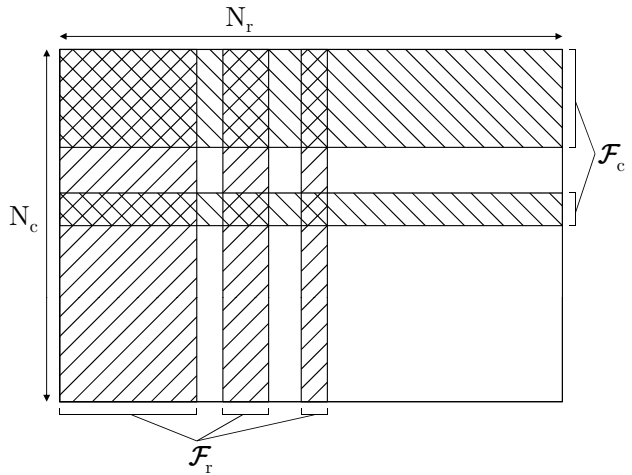}
  \caption{Input matrix $U$ for a product polar code.}
  \label{fig:prod_pol_des}
\end{figure}
As we have seen in the previous section, both polar and product codes can be defined through the Kronecker product of short and simple blocks, that are used to construct longer and more powerful codes. 
Even if systematic polar codes have been used in the construction of product codes \cite{par_conc_sys_pol}, this peculiar structure has never been really exploited in the product code construction. 
In the following, we show how to construct product codes on the basis of two non-systematic polar codes, proving that the result is again a polar code having a particular frozen set, that can be calculated on the basis of the frozen sets of the component polar codes. 
Next, we show that a polar code can be seen as an irregular product code \cite{irr_prod}, and we describe how to calculate the frozen sets of the component polar codes. 
Proofs of the Propositions can be found in the Appendix. 
The proposed design can be extended to multi-dimensional product codes. 

%\section{From Product Polar Codes to Polar Codes} \label{sec:PPC2PC}
\subsection{Product Polar Codes} \label{sec:PPC2PC}
\begin{figure}
  \centering
  \includegraphics[width=0.25\textwidth]{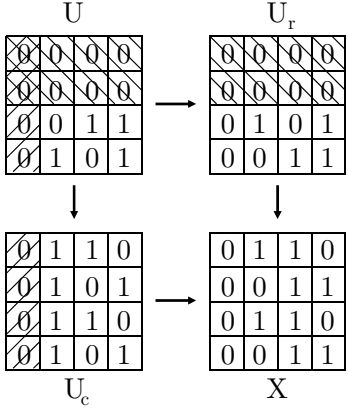}
  \caption{Example of product polar code design and encoding.}
  \label{fig:prod_pol_ex}
\end{figure}
Product polar code design starts from two polar codes $\mathcal{C}_r$ and $\mathcal{C}_c$ with parameters $(N_r,K_r)$ and $(N_c,K_c)$, having transformation matrices $T_{N_r}$ and $T_{N_c}$  and frozen sets $\mathcal{F}_r$ and $\mathcal{F}_c$ respectively. 
Encoding of product polar code $\mathcal{P} = \mathcal{C}_c \times \mathcal{C}_r$ is performed on the basis of an $N_c \times N_r$ input matrix $U$ having zeros in the rows listed in $\mathcal{F}_c$ and in the columns listed in $\mathcal{F}_r$, as depicted in Figure~\ref{fig:prod_pol_des}. 
Similarly to product codes, input bits are inserted row-by-row in the remaining $K_r K_c$ entries of $U$, starting from the top left entry. 
Product code encoding can now be performed, namely by encoding rows of $U$ using polar code $\mathcal{C}_r$ by multiplying them by the transformation matrix $T_{N_r}$. 
Columns of resulting intermediate matrix $U_r$ are further encoded using $\mathcal{C}_c$ by multiplying them by $T_{N_c}$, obtaining codeword matrix $X$. 
Again, the encoding order can be inverted without changing the result. 
The described encoding process can be mathematically expressed as
\begin{equation}
\label{eq:prod_pol_enc}
X = T_{N_c}^T \cdot U \cdot T_{N_r}.
\end{equation}

Given the linear transformation $\row(\cdot)$ converting a matrix into a row vector by juxtaposing its rows head-to-tail, we now prove that $x = \row(X)$ is the codeword of a polar code. 

%Input and codeword matrices $U$ and $X$ have to be vectorized into row vectors $u$ and $x$ in order to show that this procedure creates a polar code. 
%Given the classical vectorization function $\vect(\cdot)$ converting matrices into column vectors, here we define an analogous linear transformation $\row(\cdot)$ converting a matrix into a row vector by juxtaposing its rows head-to-tail. 
%Equipped with this definition, we can extend a classical result of $\vect(\cdot)$ function to $\row(\cdot)$ function. 
%\begin{lemma}
%\label{prop:row}
%Given three matrices $A$, $B$, $C$, if $A \cdot B \cdot C$ is defined, then
%\begin{equation}
%\row(A \cdot B \cdot C) = \row(B) \cdot (A^T \otimes C).
%\end{equation}
%\begin{proof}
%The compatibility of vectorization with the Kronecker product is a well known result, that is used to express matrix multiplication $A \cdot B \cdot C$ as a linear transformation $\vect(A \cdot B \cdot C) = (C^T \otimes A) \cdot \vect(B)$. 
%Having $\vect(A^T) = (\row(A))^T$ by construction, then
%\begin{align*}
%\row(A \cdot B \cdot C) & =  (\vect((A \cdot B \cdot C)^T))^T  \\
%  & =  (\vect(C^T \cdot B ^T\cdot A^T))^T  \\
%  & =  ((A \otimes C^T) \cdot \vect(B^T))^T  \\
%  & =  (\vect(B^T))^T \cdot  (A \otimes C^T)^T  \\
%  & =  \row(B) \cdot (A^T \otimes C).
%\end{align*}
%\end{proof}
%\end{lemma}
%Lemma~\ref{prop:row} is used to prove the main proposition of this section: 
\begin{proposition}
\label{prop:frozen}
The $(N,K)$ product code $\mathcal{P}$ defined by the product of two polar codes as $\mathcal{P} = \mathcal{C}_c \times \mathcal{C}_r$ is a polar code having transformation matrix $T_N = T_{N_c} \otimes T_{N_r}$ and frozen set %$\mathcal{F}$ given by
%polar code produced by the product of two polar codes $\mathcal{C}_c$ and $\mathcal{C}_r$ is equal to 
%with frozen sets $\mathcal{F}_c$ and $\mathcal{F}_r$ is equal to
%\[ \mathcal{F} = \mathcal{F}_c \cup \mathcal{F}_r \]
\begin{equation}
\label{eq:frozen}
\mathcal{F} = \arg \min (z_c \otimes z_r) ,
\end{equation} 
where $z_r$ ($z_c$) is a vector of length $N_r$ ($N_c$) having zeros in the positions listed in $\mathcal{F}_r$ ($\mathcal{F}_c$) and ones elsewhere. 
%\begin{proof}
%If we define $u = \row(U)$, then input vector $u$ has frozen bits imposed by \eqref{eq:frozen} according to the definition of input matrix $U$. 
%With slight abuse of notation, we use the $\arg \min$ function to return the set of the indices of vector $i = i_c \otimes i_r$ for which the entry is zero. 
%Polar codeword $x$ is calculated through Lemma~\ref{prop:row} as 
%\begin{align*}
%x & = \row(X) \\
%  & = \row(T_{N_c}^T \cdot U \cdot T_{N_r}) \\
%  & = \row(U) \cdot (T_{N_c} \otimes T_{N_r}) \\
%  & = u \cdot T_N. 
%\end{align*}
%Finally, if $N_r = 2^{n_r}$ and $N_c = 2^{n_c}$, then $T_N = T_{N_c} \otimes T_{N_r} = T_2^{\otimes(n_c + n_r)}$, hence $T_N$ is the transformation matrix of a polar code of length $N = 2^{n_c + n_r}$.
%\end{proof}
\end{proposition}

Proposition~\ref{prop:frozen} shows that the product of two polar codes is still a polar code, providing its transformation matrix and frozen set on the basis of the two component polar codes. 
The resulting product polar code $\mathcal{P}$ has parameters $(N,K)$, with $N = N_r N_c$ and $K = K_r K_c$, and frozen set $\mathcal{F}$ designed according to \eqref{eq:frozen}. 
It is worth noting that such a frozen set is suboptimal, with respect to SC decoding, compared to the one calculated for a polar code of length $N$, i.e. it does not collect the $N-K$ less reliable positions. 
On the other hand, we will see that the latency gain allowed by the product polar structure compensates the decoding performance loss. 
%The sub-vectors $x_r^i$ and $x_c^j$ corresponding to the $i$-th row and the $j$-th column of $X$ represent codewords of polar codes $\mathcal{C}_r$ and $\mathcal{C}_c$ respectively. 

Figure~\ref{fig:prod_pol_ex} shows the encoding of a product polar code generated by a $(4,2)$ polar code with frozen set $\mathcal{F}_c = \{ 0,1 \}$ as column code $\mathcal{C}_c$ and a $(4,3)$ polar code with frozen set $\mathcal{F}_r = \{ 0 \}$ as row code $\mathcal{C}_r$. 
This defines a product polar code $\mathcal{P}$ with $N = 16$ and $K = 6$. 
According to Proposition~\ref{prop:frozen}, its frozen set can be calculated through the Kronecker product of the auxiliary vectors $z_c = [0,0,1,1]$ and $z_r = [0,1,1,1]$, from which $z = [0,0,0,0,0,0,0,0,0,1,1,1,0,1,1,1]$ and $\mathcal{F} = \{ 0,1,2,3,4,5,6,7,8,12 \}$. 
We recall that the optimal frozen set for a $(16,6)$ polar code would be given by $\mathcal{F}' = \{ 0,1,2,3,4,5,6,8,9,10 \}$.

%\section{From Polar Codes to Product Polar Codes} \label{sec:PC2PPC}
%\section{Product Codes description of Polar Codes} \label{sec:PC2PPC}
\subsection{Polar Codes as Product Codes} \label{sec:PC2PPC}
\begin{figure}
  \centering
  \includegraphics[width=0.25\textwidth]{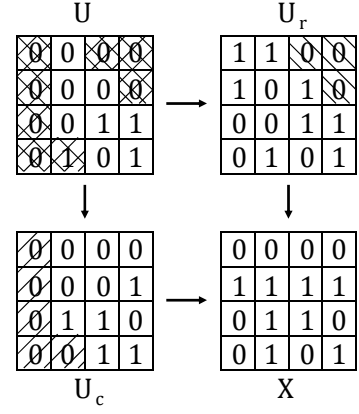}
  \caption{Example of product encoding of a polar code.}
  \label{fig:pol2prod_ex}
\end{figure}
%
%In the previous Section, we have shown how the product of two polar codes is a polar code. 
Polar codes are inherently recursive in nature, being defined through the $n$-fold Kronecker product of the polarization kernel $T_2$. 
%given that their transformation matrix $T_N$ is the $n^{\rm th}$ Kronecker power of the polarization kernel. 
%It is thus obvious that a polar code can be seen as the product of two or more shorter polar codes.
This structure makes it possible to separate the blocks composing the polar code, describing it as an irregular product code~\cite{irr_prod}, i.e. a product code composed by codes of different rates in the same encoding direction. 

Let polar code $\mathcal{P}$ be defined by the transformation matrix $T_N = T_2^{\otimes n}$ and the frozen set $\mathcal{F}$. 
Let us rearrange the codeword $x = u \cdot T_N$ and the input vector $u$ in two $N_c \times N_r$ matrices $X$ and $U$ row-by-row; according to \eqref{eq:prod_pol_enc}, $X$ can be obtained by $U$ through product code encoding. 
As a consequence, every row (column) of $X$ can be seen as codeword of a polar code of length $N_r$ ($N_c$). 
%However, the frozen sets of these polar codes are not given; to be able to decode the row (column) codewords, it is necessary to infer their frozen set from $\mathcal{F}$. 
The knowledge of the frozen sets of these codes is necessary to run the product decoding algorithm, however they are not given in the construction; Proposition~\ref{prop:frozen_small} will show how to infer them from $\mathcal{F}$. 
%The frozen sets of these polar codes are not given; to be able to decode the row (column) codewords, it is necessary to infer their frozen set 
Given the length $N$ vector $z$ having zeros in the positions listed in the frozen set $\mathcal{F}$ and ones elsewhere, the $N_c \times N_r$ matrix $Z$ is defined by reshaping $z$ row-by-row. %$Z$ is used in the following proposition to define the component polar codes of the irregular product code.  
In the following, $A^{(i,\cdot)}$ and $A^{(\cdot,j)}$ represent the $i$-th row and the $j$-th column of matrix $A$ respectively. 
%; however, what is the rate of this polar code?
%To be able to decode the row (column) codewords, we need to infer their frozen set from $\mathcal{F}$.
%
%If the rate is 1, the above sentence is obviously true; however, is it possible to have polar codes with different rates?
%To answer to this question, we need the frozen set of every row and column code. 

\begin{proposition}
\label{prop:frozen_small}
For a given polar code $\mathcal{C}$, the frozen sets $\mathcal{F}_r^i$ and $\mathcal{F}_c^j$ of its $i$-th row component polar code and $j$-th column component polar code are given by
\begin{equation}
\mathcal{F}_c^j = \arg \min \left(Z_r^{(\cdot,j)}\right) \text{ and }
\mathcal{F}_r^i = \arg \min \left(Z_c^{(i,\cdot)}\right) , \label{eq:frozen_small}
\end{equation}
%\begin{eqnarray}
%\mathcal{F}_c^j = \arg \min \left(Z_r^{(\cdot,j)}\right) \label{eq:frozen_small_c} \\
%\mathcal{F}_r^i = \arg \min \left(Z_c^{(i,\cdot)}\right) , \label{eq:frozen_small_r}
%\end{eqnarray}
where $Z_r = Z \ast T_{N_r}$, $Z_c = T_{N_c}^T \ast Z$ and the operator "$\ast$" represents multiplication over $\mathbb{N}$. 
As before, $\arg \min$ is used to extract the indices of the zero entries of its operand. 
%where $z_r$ ($z_c$) is a vector of length $N_r$ ($N_c$) having zeros in the positions listed in $\mathcal{F}_r$ ($\mathcal{F}_c$) and ones elsewhere. 
\end{proposition}

Proposition \ref{prop:frozen_small} permits to describe a polar code as an irregular product code \cite{irr_prod}, namely a product code for which every row and column is defined by a different polar code. 
To sum up, a polar code $\mathcal{P}$ with transformation matrix $T_N$ and frozen set $\mathcal{F}$ can be described as an irregular $N_c \times N_r$ product code, where the $i$-th row ($j$-th column) component code is a polar code $\mathcal{C}_r^i$ ($\mathcal{C}_c^j$) of length $N_r = 2^{n_r}$ ($N_c = 2^{n_c}$) with transformation matrix $T_{N_r}$ ($T_{N_c}$) and frozen set $\mathcal{F}_r^i$ ($\mathcal{F}_c^j$) defined by Proposition~\ref{prop:frozen_small}. 
%Proposition~\ref{prop:frozen_small_r} (Proposition~\ref{prop:frozen_small_c}). 
It is worth noting that the component code lengths $N_r$ and $N_c$ are not defined, and they can assume any value provided that their product matches the polar code length $N$. 
Different component code lengths provide different component code dimensions, hence these two parameters should be carefully chosen to limit the number of rate-one component codes. 

The average $R_r$ and $R_c$ thus obtained are higher than what would impose a polar code rate $R$ in the construction detailed in Section \ref{sec:PPC2PC}, as many frozen bits in $\mathcal{F}$ impose row and column codeword constraints that are not reflected in $\mathcal{F}_r$ and $\mathcal{F}_c$, and are consequently not exploited in SC-based decoding. 
For this reason, we will see that product decoding of classical polar codes have poor error-correction performance compared to plain SC decoding.
%We will propose a frozen set design strategy to improve product decoding of polar codes in Section~\ref{sec:frozenSetDesign}. 
We propose a frozen set design improving product decoding of polar codes in next section. 

As an example, let us take a $(16,8)$ polar code $\mathcal{P}$ with frozen set $\mathcal{F} = \{ 0,2,3,4,7,8,12,13 \}$. 
If $N_r = N_c = 4$, then
\begin{equation}
Z = \left[ \begin{matrix}
0 & 1 & 0 & 0 \\
0 & 1 & 1 & 0 \\
0 & 1 & 1 & 1 \\
0 & 0 & 1 & 1
\end{matrix} \right] \notag
\end{equation}
and the frozen sets of component polar codes are calculated using
\begin{equation}
\small
%\tiny
Z_r = Z \ast T_4 = \left[ \begin{matrix}
1 & 1 & 0 & 0 \\
2 & 1 & 1 & 0 \\
3 & 2 & 2 & 1 \\
2 & 1 & 2 & 1
\end{matrix} \right]
Z_c = T_4^T \ast Z = \left[ \begin{matrix}
0 & 3 & 3 & 2 \\
0 & 1 & 2 & 1 \\
0 & 1 & 2 & 2 \\
0 & 0 & 1 & 1
\end{matrix} \right] \notag
\end{equation}
Row polar codes have frozen sets defined through the rows of $Z_c$, with $\mathcal{F}_r^0 = \mathcal{F}_r^1 = \mathcal{F}_r^2 = \{ 0 \}$ and $\mathcal{F}_r^3 = \{ 0,1 \}$, while columns polar codes have frozen sets defined through the columns of $Z_r$ as $\mathcal{F}_c^0 = \mathcal{F}_c^1 = \emptyset$, $\mathcal{F}_c^2 = \{ 0 \}$ and $\mathcal{F}_c^3 = \{ 0,1 \}$. 
Row polar codes have then dimension $(3,3,3,2)$, while column polar codes have dimension $(4,4,3,2)$. 
Product encoding of this polar code $\mathcal{P}$ is depicted in Figure~\ref{fig:pol2prod_ex}. 
%In this case, the rows are polar codes of length $N_r = 4$ with frozen sets, according to Proposition \ref{prop:frozen_small_r}, $\mathcal{F}_r^0 = \mathcal{F}_r^1 = \mathcal{F}_r^2 = \{ 0 \}$ and $\mathcal{F}_r^3 = \{ 0,1 \}$, while the columns are polar codes of length $N_c = 4$ with frozen sets, according to Proposition \ref{prop:frozen_small_c}, $\mathcal{F}_c^0 = \mathcal{F}_c^1 = \emptyset$, $\mathcal{F}_c^2 = \{ 0 \}$ and $\mathcal{F}_c^3 = \{ 0,1 \}$. 

%\section{Mixed frozen set design for two-step decoding of polar codes} \label{sec:frozenSetDesign}
%\section{Frozen set design for two-step decoding} \label{sec:frozenSetDesign}
\subsection{Hybrid frozen set design for product decoding} \label{sec:frozenSetDesign}

%In Section~\ref{sec:PC2PPC} we have seen that the frozen set $\mathcal{F}$ of a polar code infers the frozen sets of the component polar codes in the irregular product code construction. 
The selection of $\mathcal{F}$ according to bit channel polarization \cite{arikan} may result in very inefficient component polar codes, since some of them may include very few frozen bits. 
%As a result, the first step of the proposed decoding approach may be ineffective, leading to $\gamma\approx 1$ and longer decoding latency. 
%However, this code construction leads to better error-correction performance, since it maximizes the effectiveness of SC decoding of the overall polar code.
As a result, the product code decoding approach may be ineffective for classical polar codes, even if this code construction leads to better error-correction performance under full SC decoding. %, since it maximizes the effectiveness of SC decoding of the overall polar code. 
On the other hand, product polar code design, imposing $\mathcal{F}$ on the basis of component polar codes, leads to a suboptimal frozen set for the full code and then to worse error-correction performance under full SC decoding. 
%On the other hand, product polar code designs, that impose an optimal frozen set on $\mathcal{F}_r$ and $\mathcal{F}_c$ and infer $\mathcal{F}$ from them, lead to short latency but worse error-correction performance. 
%This is due to the fact that the effectiveness of first decoding step is maximized, sharply decreasing $\gamma$. 
%However, in case product decoding fails, the polar decoding approach has to rely on a suboptimal frozen set $\mathcal{F}$. 
%Since product decoding is generally weaker than polar decoding, the resulting error-correction performance is worse. 
To overcome these problems, we propose an ad-hoc frozen set selection to find a trade-off between error-correction performance and decoding latency. 
Our goal is to propose a frozen set including the majority of degraded bit channels, to exhibit adequate error-correction performance, however maintaining well designed component polar codes.  

%The error-correction performance and the decoding latency can be tuned by applying an ad-hoc frozen set selection. 
Let us define as $R$ the desired rate of the length-$N$ polar code, and as $R_r$ and $R_c$ the rates of the row and column component codes, where $R_r\cdot R_c > R$. 
As a first step, $\mathcal{F}_r$ and $\mathcal{F}_c$ are designed targeting optimal SC-based decoding of length-$N_r$ and length-$N_c$ polar codes, as for product polar codes design. 
In this way, a frozen set $\mathcal{F}$ is inferred to the length-$N$ polar codes, having rate $R_r\cdot R_c$. 
%The inferred $\mathcal{F}$ has rate $R_r R_c$: i
Since $R_r\cdot R_c> R$, the remainder of the frozen bit positions needed to achieve $R$ is set as the least reliable positions of the length-$N$ polar code that are not already frozen in $\mathcal{F}$.
In practice, the difference between $R_r\cdot R_c$ and $R$ allows to trade-off latency and performance. 
%As the difference increases, $\gamma$ tends to zero with a gentler slope, and the effectiveness of the second decoding step increases, leading to better error-correction performance. 
%With $R_r\times R_c = 1$, the frozen set $\mathcal{F}$ is the optimal one for polar decoding, while if $R_r\times R_c=R$, the frozen set $\mathcal{F}$ is the one inferred by $\mathcal{F}_r$ and $\mathcal{F}_c$.
%
This construction approach can lead to undetected errors under product decoding, as both row and column decoding phases can agree on a candidate codeword that does not take in account the additional frozen bits in $\mathcal{F}$. 
This effect can be limited by re-encoding the codeword identified in the first decoding phase and checking if the bits in $\mathcal{F}$ have all been assigned a $0$. 
%In case not, the second decoding step is activated.

\section{Two-step decoding of polar codes} \label{sec:dec}

\begin{algorithm}[t!]
\caption{TwoStepDecoding} \label{alg:2step}
\begin{algorithmic}[1]
%\Procedure{MyProcedure}{}
\STATE Initialize $Y_r = Y_c = Y $
\FOR{$i = 1 \dots t$}
   \STATE $\hat{U}_c = \text{DecodeRows}(Y)$
   \STATE $\hat{U}_r = \text{DecodeCols}(Y)$
   \STATE $\hat{X}_c = \hat{U}_c \cdot T_{N_r}$
   \STATE $\hat{X}_r = T_{N_c}^T \cdot \hat{U}_r$
   \IF{$\hat{X}_r == \hat{X}_c$}
%      \STATE $\hat{x} = \row(X_r)$
%      \RETURN $\hat{u} = \text{PolarEncoding}(\hat{x})$ \label{alg:step:invol}
      \RETURN $\hat{u} = \text{PolarEncoding}(\row(\hat{X}_r))$ \label{alg:step:invol}
%      \STATE break
   \ELSE
      \STATE $(Y_r,Y_c) = \text{UpdateLLRs}$
   \ENDIF 
\ENDFOR
\RETURN $\hat{u} = \text{Decode}(\row(Y))$  
%\EndProcedure
\end{algorithmic}
\end{algorithm}

%Proposed decoding approaches for low-latency/low-complexity product-like decoding.
In this Section, we present a two-step decoding scheme for polar codes, based on their interpretation as both polar codes and product codes. 
This dual nature permits to initially decode the code as a product code (step 1), and in case of failure to perform polar decoding on the full polar code (step 2). 
%We propose to initially decode the code as a product code (step 1), and in case of failure to perform polar decoding on the full polar code (step 2). 
During step 1, row and column decoders can exchange either hard or soft decisions, while step 2 is always performed by a SC-based decoder. 
In the following, we detail several incarnations of this two-step decoding approach.

The first decoding step considers the polar code as a product code.
Vector $y$ containing the log-likelihood ratios (LLRs) of the $N$ received bits is rearranged in the $N_c \times N_r$ matrix $Y$ row-by-row.  
Every row (column) is considered as a noisy $\mathcal{C}_r$ ($\mathcal{C}_c$) polar codeword, and decoded independently. 
The row and column decodings might occur at the same time or one after the other, and they might exchange soft or hard information, and might be repeated for $t$ iterations or until a stopping criterion is met. 
In case residual errors are detected at the end of this first decoding step, a second decoding step is performed by decoding the code as a full polar code. 
The proposed decoding technique is summarized in Algorithm~\ref{alg:2step}; it is worth noticing that, due to involution property of the transformation matrix of polar codes, hard decoding of estimated codeword $\hat{x}$ at line \ref{alg:step:invol} can be performed through polar encoding. 
%The decoding algorithm used to decode the component codes and the full polar code, along with the information exchange technique between row and column decoding phases during step 1 and their scheduling, are design choices yielding different trade-offs between error correction performance and complexity. 
The decoding algorithm employed in the two decoding steps, along with the information exchange technique between row and column decoding phases during step 1 and their scheduling, are design choices yielding different trade-offs between error correction performance and complexity. 

The structure of parallel and partially-parallel SC-based decoders is based on a number of processing elements performing LLR and hard decision updates, and on dedicated memory structures to store final and intermediate values. 
Given the recursive structure of polar codes, decoders for shorter codes are naturally nested within decoders for longer codes. 
In the same way, the main difference between long and short code decoders is the amount of memory used. 
Thus, regardless of the chosen decoding algorithm, not only a high degree of resource sharing can be expected between the first and second decoding step; the parallelization available during the first decoding step implies that the same hardware can be used in the second step, with minor overhead.

%\subsection{Comparison of Hard Decisions} \label{subsec:dec}
\subsection{Hard Decision (HD) Decoding} \label{subsec:dec}

\begin{figure}
  \centering
  \includegraphics[width=0.45\textwidth]{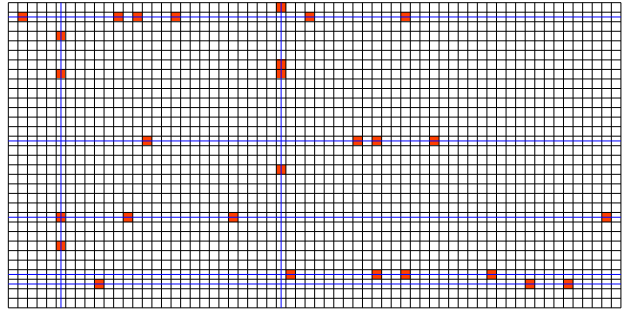}
  %\caption{Example of overlapping of $\hat{X}_r$ and $\hat{X}_c$. 
  \caption{Example of $X_d$ estimate; red squares represent mismatches, blue lines represent errors located by Algorithm \ref{alg:find_err}.}
  \label{fig:err_patt}
\end{figure}
SC is a soft-input / hard-output decoding algorithm.
%By using SC-like algorithms to decode the component codes and the full polar codes, and exchanging their hard output between the row and column decoding phases, we can obtain a low-complexity hard decision decoder as follows. 
Leveraging on this property, at step 1 we propose to decode the component codes through an SC-based decoder and exchange their hard output between the row and column decoders. 
In this way, we obtain a low-complexity hard decision decoder for product polar codes as follows. 

To begin with, every row of $Y$ is decoded through the SC-based algorithm to obtain the estimate binary matrix $\hat{U}_c$ 
Each row of $\hat{U}_c$ is re-encoded independently, obtaining $\hat{X}_c=\hat{U}_c\cdot T_{N_r}$. %: the $N_r$-bit vectors $\hat{x}_r$ are then stored as rows of matrix $X_r$. 
The same procedure is applied at the same time to the columns of $Y$, obtaining estimated matrix $\hat{U}_r$ that is used to estimate codeword matrix as $\hat{X}_r = T_{N_c}^T \cdot \hat{U}_r$. % $\hat{x}_c=\hat{u}_c\cdot T_{N_c}$, that are in turn stored as columns of matrix $X_c$. 
%The exchange of information between the two stages is performed by comparing $\hat{X}_r$ to $\hat{X}_c$; i
If $\hat{X}_r = \hat{X}_c$, decoding is considered successful and the estimated input vector $\hat{u}$ of code $\mathcal{P}$ can thus be derived by encoding vector $\hat{x}=\row(\hat{X}_r)$, since $T_N$ is involutory. %, i.e. via matrix multiplication with the transformation matrix $T_N$. 
%matrix multiplication of the vectorized version of $X_r$ with the transformation matrix $T_N$. 
In case $\hat{X}_r \neq \hat{X}_c$, soft inputs have to be updated by exchanging hard information between rows and columns. 

%In this paper we propose to estimate erroneous row and column codewords comparing row and columns estimates and update their LLRs on the basis of the codebits estimated in the other direction. 
%We propose to update LLRs on the basis of a comparison between $\hat{X}_r$ and $\hat{X}_c$. 
We propose to update LLRs on the basis of an estimation of the error committed by row and column decoders. 
Incorrect rows can be rectified using correct columns by saturating the corresponding LLR and vice-versa; however, LLRs of intersections of wrong rows and columns cannot be updated in this way.
In order to correct these errors, we propose to treat the intersection points as erasures by zeroing their LLRs.  
As an example, in a row, crossing points with incorrect columns have their LLR set to 0, while intersections with correct columns set the LLR to $+\infty$ if the bit occupying the same position in $\hat{X}_c$ has been decoded as $0$, and to $-\infty$ if the bit is a $1$. 
After the update, another row and column decoding step is performed; it is worth noticing that only rows and columns flagged as incorrect need to be re-decoded. %, obtaining updated $\hat{X}_r$ and $\hat{X}_c$.
This procedure is iterated a number $t$ of times, or until $\hat{X}_r = \hat{X}_c$.
%In case $X_r \neq X_c$ after $t$ iterations, the second step of the proposed decoding approach is activated. 
%The algorithm used to decode the component codes in step one is used to decode the received vector $y$, considering the complete code $\mathcal{P}$. 
If $\hat{X}_r \neq \hat{X}_c$ after $t$ iterations, the first step returns a failure.
In this case, the second step of the algorithm is performed, namely the received vector $y$ is decoded directly, considering the complete length-$N$ polar code $\mathcal{P}$.

\begin{algorithm}[t!]
\caption{FindErroneousEstimations} \label{alg:find_err}
\begin{algorithmic}[1]
%\Procedure{MyProcedure}{}
\STATE Initialize $\text{ErrRows} = \text{ErrCols} = \emptyset$
\STATE $X_d = \hat{X}_r \oplus \hat{X}_c$%(X_r \neq X_c)$
\STATE $\text{NumErrRows} = \text{SumRows}(X_d) $
\STATE $\text{NumErrCols} = \text{SumCols}(X_d) $
\WHILE{$\text{NumErrRows} + \text{NumErrCols} > 0$}
	  \STATE $e_r=\text{arg max(NumErrRows)}$ 
  	  \STATE $e_c=\text{arg max(NumErrCols)}$ 
   \IF{$\text{max(NumErrRows)} > \text{max(NumErrCols)}$}
      \STATE $\text{ErrRows} = \text{ErrRows}  \cup \{e_r\}$
      \STATE $X_d(e_r,:) = 0$
   \ELSE
      \STATE $\text{ErrCols} = \text{ErrCols} \cup \{e_c\}$
      \STATE $X_d(:,e_c) = 0$
   \ENDIF
   \STATE $\text{NumErrRows} = \text{SumRows}(X_d) $
   \STATE $\text{NumErrCols} = \text{SumCols}(X_d) $
\ENDWHILE
\RETURN ErrRows, ErrCols
%\EndProcedure
\end{algorithmic}
\end{algorithm}

Incorrect rows and columns can be identified studying the pattern of mismatches of matrix $X_d = \hat{X}_r \oplus \hat{X}_c$ having ones in the positions where the row and column decoders disagree.
%it is possible to identify incorrect estimations by overlapping $\hat{X}_r$ and $\hat{X}_c$ and observing the pattern of mismatches.
%Since decoding errors in a row are usually correct in the intersecting columns, and vice-versa, m
Mismatches are usually grouped in strings, as shown in Figure~\ref{fig:err_patt}, where they are represented by red squares. 
% then an error occurred in the decoding. 
%This means that either some row or some column have been decoded incorrectly. 
%Overlapping $X_r$ and $X_c$, it is possible to guess these incorrect codewords since differences between the two matrices are grouped in strings, as shown in Figure~\ref{fig:err_patt}. 
%In general, the lower the component codes rate, the higher is the number of mismatches of an incorrect codeword. 
Even if mismatch patterns are simple to analyze by visual inspection, it may be complex for an algorithm to recognize an erroneous row or column. 
Therefore we propose a greedy algorithm to accomplish this task. 
%detect the incorrect codewords (\textbf{describe algo}) . 
In the proposed method, described as Algorithm~\ref{alg:find_err}, the number of mismatches in each row and column is initially counted, and the row or column with the highest count is flagged as incorrect. %flagging as incorrect that with the highest count.
Next, its contribution is subtracted from the mismatch count of connected rows or columns, and another incorrect row or column is identified. 
The process is repeated until all mismatches belong to at least one incorrect row or column. %; their indices are stored in $\text{ErrRows}$ and $\text{ErrCols}$ respectively.
An example of this identification process is represented by the blue lines in Figure~\ref{fig:err_patt}.

\subsection{Soft Decision (SD) Decoding}\label{subsec:soft}

%%
%\begin{algorithm}[t!]
%\caption{SDUpdateLLRs} \label{alg:SDcomp}
%\begin{algorithmic}[1]
%%\Procedure{MyProcedure}{}
%\STATE ...
%\RETURN $(Y_r,Y_c)$  
%%\EndProcedure
%\end{algorithmic}
%\end{algorithm}

Product code decoding algorithms have long benefited from the exchange of soft information between row and column decoding phases \cite{CHASE}. 
In our soft decision decoder, each row of $Y$ is initially decoded through a soft-output decoder, obtaining new soft values for the received symbols that are stored in $\widetilde{X}_c$. 
The same procedure is applied simultaneously to the columns of $Y$, obtaining estimated codeword matrix $\widetilde{X}_r$. 
If $\sgn \left( \widetilde{X}_r \right) = \sgn \left( \widetilde{X}_c \right)$, decoding is considered successful and the estimated input vector $\hat{u}$ can be derived, otherwise soft information has to be exchanged between rows and columns to continue the decoding. 
%The soft information is then taken as an input by the following decoding phase.
Soft information calculated by row decoders is provided as input for the column soft decoder, and vice versa. 
As with hard decoding, a maximum number $t$ of iterations is performed before going to second decoding step. 
%the first decoding step is stopped as soon as $X_r=X_c$, or if the maximum number of iterations $t$ has been reached. 

While SC-based decoding algorithms are inherently soft-input/hard-output, BP has been used in polar code decoding \cite{BP_pc}, and SCAN has been proposed in \cite{SCAN_pc} as a soft-output version of SC. 
These algorithms however rely on multiple iterations to refine their soft information and improve or even reach the error-correction performance of SC-based algorithms. 
Since product decoding is an iterative process itself, an iterative component decoding might lead to very large decoding latency. 

%We considered both these algorithms for use in our two-step decoding approach, as component code decoders for the first decoding step and as post-processing for the second step. 
%Their soft output has been used as an updated LLR passed between the row and column decoding phases during the first step. 
%These algorithms, however, rely on a relatively high number of iterations to refine the soft information, while their error-correction performance is poor at a low iteration count \cite{SCAN_pc}.
%Moreover, short codes adversely affect the algorithm performance with respect to SC-based decoding algorithms.
%Nevertheless, simulation results obtained with both BP and SCAN have shown that, even with a very high number of component decoder iterations, the performance of the first decoding step is very poor, and the second step is activated with high probability. }{Revise}

%\fixme{Cite and comment Soft SCL}{}

Inspired by the Chase decoding principle \cite{CHASE}, we propose an alternative way to obtain soft information from the different decoding candidates available through list decoding. 
Let us consider the LLR-based formulation of SCL in \cite{balatsoukas_SCL_HW}, where to each candidate paths $\hat{u}^{(l)}$ for $i=0,\dots,L-1$ is assigned a path metric $M_l$ computed as the sum of the LLRs for which the estimated bit is not equal to the hard decision; a path metric can be hence calculated as 
\begin{equation}
\label{eq7}
M_l = \sum_{i=0}^{N-1} a_i^{(l)} \cdot \alpha_i^{(l)}
\end{equation}
where $\alpha_{i}^{(l)}$ is the LLR associated to bit $\hat{u}_{i}^{(l)}$ and $a_i^{(l)} = \hat{u}_{i}^{(l)} - \left(1-\sgn \left(\alpha_i^{(l)} \right) \right)/2$. 
%$\PM$, updated after each bit estimation $\hat{u}_{i}$ as
%\begin{align}
%\PM_{{i}} = \begin{cases}
%    \PM_{{i-1}} + |\alpha_{i}| \text{,} & \text{if } \hat{u}_{i} \neq \text{HD}(\alpha_{i})\text{,}\\
%    \PM_{{i-1}} \text{,} & \text{otherwise,}
%  \end{cases} \label{eq7}
%\end{align}
%where $\alpha_{i}$ is the LLR associated to $\hat{u}_{i}$, $\PM_0 = 0$ and
%\begin{align}
%\text{HD}(\alpha_{i}) = \begin{cases}
%    0 & \text{if } \alpha_{i}\ge 0\text{,}\\
%    1 & \text{otherwise.}
%  \end{cases}
%\end{align}
At the end of the SCL decoding, we take the $L$ estimated input vectors  $\hat{u}^{(0)},\dots,\hat{u}^{(L-1)}$, having path metrics $M_0,\dots,M_{L-1}$, and re-encode them obtaining the estimated codewords $\hat{x}^{(0)},\dots,\hat{x}^{(L-1)}$. 
Soft information $\Lambda_i$ associated to code bit $x_i$ is then calculated as
\begin{equation} \label{eq:soft}
\Lambda_i=\min_{\hat{x}^{(l)}_i = 1}(M_l)-\min_{\hat{x}^{(l)}_i = 0}(M_l)~,
\end{equation}
namely as the path metric difference between the most reliable codewords assigning 1 and 0 to code bit $x_i$. 
In case all codewords have the same value for a given bit $x_i$, a large value is assigned to $\Lambda_i$, signifying the agreement of all candidates.
%i.e. $\{ l=0,\dots,L-1 \, s.t. \, \hat{x}^l_i = a\} = \emptyset$, $\Lambda_i$ is assigned a large value, signifying the agreement of all candidates.
% 
%Let us take the $L$ estimated vectors ${\bf \hat{u}_{l}}$ and re-encode them to obtain the estimated codewords ${\bf \hat{s}_{l}}$. 
%To compute the soft information $\Lambda_j$ associated to each codeword bit $j$, it is sufficient to compute
%\begin{equation} \label{eq:soft}
%\Lambda_j=\min(\PM^1)-\min(\PM^0)~,
%\end{equation}
%where $\PM^1$ is the set of $\PM$s associated to the candidate codewords for which bit $j$ is equal to $1$. 
%In the same way, $\PM^0$ is the set of $\PM$s associated to the candidate codewords for which bit $j$ is equal to $0$.
%It is shown in (\ref{eq7}) that the path metric is computed as the sum of the LLRs for which the estimated bit is not equal to the hard decision. 
%It is thus a measure of the discrepancy between the received vector and the estimated one, and can be viewed as the cost of decoding to $\hat{u}$ given the received vector $y$. 
Path metric is in practice a measure of the discrepancy between the received vector and the estimated one, and can be viewed as the cost of decoding to $\hat{u}$ given the received vector $y$. 
In (\ref{eq:soft}), one of the two members of the equation is always the minimum among all $M$, i.e. the path metric associated to the path chosen as the result of the decoding process. 
Equation (\ref{eq:soft}) thus expresses the reliability of each bit estimation, as the bit-by-bit difference between the lowest decoding cost and its closest competitor. 
Its reliance on $M$, that is available at the end of SCL decoding, makes it a low-complexity option to obtain soft information.

%The soft information is then taken as an input by the following decoding phase.
%As with hard decoding, the first decoding step is stopped as soon as $X_r=X_c$, or if the maximum number of iterations $t$ has been reached. 
%The proposed soft decoder is detailed in Algorithm~\ref{alg:SDcomp}.

%\fixme{Check if $\Lambda^{new}_{in}=\alpha(\Lambda_{out}-\Lambda^{old}_{in})+\Lambda^{old}_{in}$ gives better performance}{}

\subsection{Decoding Latency Analysis} \label{subsec:latency}

The proposed two-step decoding of product polar codes allows to split the polar decoding process into $N_r+N_c$ shorter, independent decoding processes, using the long polar code decoding only for the case of case of failure.
Let us define as $\delta_N$ the number of time steps required by a decoding algorithm to decode a polar code of length $N$. 
For the purpose of latency analysis, we suppose the decoder to have unlimited computational resources, allowing a fully parallel implementation of decoding algorithms. 

%Using Algorithm~\ref{alg:HDcomp} 
Using the proposed hard decision decoder for component codes, the expected number of steps for the proposed two-step decoder for a code of length $N = N_c\cdot N_r$ is given by
%Also, we identify as P-SC the proposed two-step decoding with component codes decoded with SC. P-SC decoding of a code of length $N=N_c\times N_r$ requires 
%\Delta^{\rm P-SC}_N = \left(2\times\max(N_r,N_c)-2\right)\times t_{avg} + \gamma(2N-2)~,
\begin{equation} \label{eq:timeHD}
\Delta^{\rm HD}_N = t_{avg} \delta_{\max(N_r,N_c)} + \gamma \delta_N~,
\end{equation}
where $t_{avg}\le t$ is the average number iterations, and $\max(N_r,N_c)$ assumes that the decoding of row and column component codes is performed at the same time.
The parameter $\gamma$ is the fraction of decoding attempts in which the second decoding step was performed. 
%It can be seen that as long as $\gamma\approx 0$ and $t_{avg}<<N/\max(N_r,N_c)$, then $\Delta^{\rm P-HD}_N$ is substantially smaller than $\Delta_N$. 
The two-step decoding latency $\Delta^{\rm HD}_N$ is substantially smaller than the full polar code decoding latency $\delta_N$ as long as $\gamma\approx 0$ and $t_{avg} \ll N/\max(N_r,N_c)$. 

In case of soft information exchange, the decoding latency can be instead computed as
\begin{equation} \label{eq:timeSD}
\Delta^{\rm SD}_N = t_{avg} ( \delta_{N_r} + \delta_{N_c} ) + \gamma \delta_N~.
\end{equation}
In this case, row and column decoding cannot be run in parallel, since the two processes need to exchange information. 
Both $t_{avg}$ and $\gamma$ are however expected to be smaller than those required in case of hard decision exchange, due to the higher quality of transferred information. 
As a consequence, soft decoding latency will be comparable to hard decision decoding when SCL is used as component decoder. 
The LLR update in (\ref{eq:soft}), while increasing the complexity of computations, has in fact negligible impact on the decoding latency when compared to SCL decoding, as all $\Lambda$ values can be calculated concurrently.

Table \ref{tab:speedHD} reports $\delta_N$ required by standard SC and SCL decoders, and the relative $\Delta_N$ for the proposed two-step decoder SC-HD, SCL-HD, and SCL-SD, at different code lengths and rates. 
For SC decoding, $\delta^{\rm SC}_N=2N-2$, while for SCL $\delta^{\rm SCL}_N=2N+K-2$ \cite{balatsoukas_SCL_HW,hashemi_FSSCL_TSP}. 
For the proposed decoding approach, $\Delta_N$ is evaluated for both hard decision (HD) and soft decision (SD) decoding, in the worst case (WC), that assumes $t_{avg}=t=4$ and $\gamma=1$, and best case (BC), that assumes $t_{avg}=1$ and $\gamma=0$.
Simulation results presented in Section \ref{sec:perf} show that $\Delta_N$ tends to the asymptotic limit represented by BC decoding latency as the channel conditions improve.

\begin{table*}
\centering
%\scriptsize
%\caption{Time step analysis for standard and two-step decoding. Worst case (WC) considers $t_{avg}=4$, $\gamma=1$. Best case (BC) considers $t_{avg}=1$, $\gamma=0$.}
\caption{Time step analysis for standard and two-step decoding.}
\label{tab:speedHD}
\setlength{\extrarowheight}{1.5pt}
\begin{tabular}{c||c|cc||c|cc|cc}

Code & \multirow{2}{*}{$\delta^{\rm SC}_N$} & \multicolumn{2}{c||}{$\Delta^{\rm HD}_N$} & \multirow{2}{*}{$\delta^{\rm SCL}_N$} & \multicolumn{2}{c|}{$\Delta^{\rm HD}_N$} & \multicolumn{2}{c}{$\Delta^{\rm SD}_N$} \\ 

$N$,$K$ & & WC & BC & & WC & BC & WC & BC\\ 
\hline
\hline
$1024, 784$  &2046 & 2294 & 62 &2830 & 3190& 90 & 3550& 180 \\ 
$1024, 841$  &2046 & 2294 & 62 &2876 & 3240& 91 & 3604& 182 \\ 
$4096, 3136$  &8190 & 8694 & 126 &11326 & 12054& 182 &12782 & 364\\ 
$4096, 3249$ &8190 & 8694 &126  &11508 & 12244& 184 &12980 & 368\\ 
$16384, 12544$ &32766 & 33782 & 254&45310  & 46774& 366 & 48238 & 732\\ 
$16384, 13225$ &32766 & 33782 &254 &46038 & 47518& 370 & 48998 & 740\\ 
$65536, 50176$ &131070 & 133110 &510 &181246  & 184182& 734 & 187118& 1468\\ 
$65536, 52900$ &131070 & 133110 &510 &184155  & 187119& 741 & 190083 & 1482\\ 
$262144, 200704$  &524286 & 528374 & 1022 &724990 & 730870& 1470 &736750 & 2940\\ 
$262144, 211600$ &524286 & 528374 & 1022 &736623 & 742555& 1483 & 748487 & 2966\\ 
\end{tabular}
\end{table*}

When code-structure-based pruning algorithms \cite{sarkis,hashemi_SSCL,hashemi_FSSCL_TSP,Condo_GLOBECOM} are used, the relationship among $\Delta_N$, $\delta_N$ and $\delta_{N_r}$ is dependent on the frozen sets of the component codes and of the length-$N$ code, and can vary from \eqref{eq:timeHD} and \eqref{eq:timeSD} significantly. 
%\fixme{Decoding complexity is comparable to hard decoding when SCL is used as polar decoder. 
%The augmented complexity due to LLRs update is in fact negligible compared to SCL decoding, since it simply represents the subtraction of two real values for every matrix $Y$ entry, and it can be calculated concurrently. }{Da rivedere a cose fatte}

\section{Performance results}\label{sec:perf}

%\fixme{Start copy pasted from \cite{PRODPOL_FIRST}. Rewrite according to the new setup as two-step decoding being only a framework.}{}

%\fixme{BER/FER and maybe also latency}{}

The dual nature of product polar codes can bring substantial speedup in the decoding; on the other hand, given a time constraint, longer codes can be decoded, leading to improved error-correction performance. 
In this Section, we present decoding speed and error-correction performance analysis, along with simulation results, for the different incarnations of the two-step decoding framework presented in Section \ref{sec:dec}.
We assume an additive white Gaussian noise (AWGN) channel with binary phase-shift keying (BPSK) modulation, while the two component codes have the same parameters, i.e. $N_r = N_c$ and $K_r = K_c$. 
The hard decision comparison incarnation of the two-step decoding framework proposed in Section \ref{subsec:dec} is labeled as SC-HD in case of SC component decoding, and SCL-HD in case of SCL decoding, while the soft-information-based decoder proposed in Section \ref{subsec:soft} is labeled as SCL-SD.
If an optimal-length CRC is concatenated to the polar code of length $N$, the second decoding step in both SCL-SD and SCL-HD can benefit from a performance improvement comparable to that observed in standard polar decoding. 
Without loss of generality, we do not consider CRC concatenation in our performance analysis.

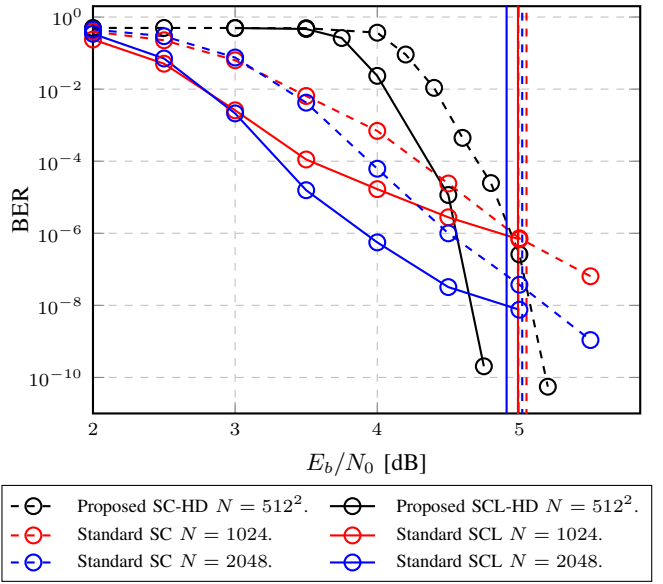
\begin{figure}
  \centering
   \scalebox{1}{\begin{tikzpicture}
  \pgfplotsset{
    label style = {font=\fontsize{9pt}{7.2}\selectfont},
    tick label style = {font=\fontsize{7pt}{7.2}\selectfont}
  }

\begin{axis}[
	scale = 1,
    ymode=log,
    xlabel={$E_b/N_0$ [\text{dB}]}, xlabel style={yshift=0.4em},
    ylabel={BER}, ylabel style={yshift=-0.75em},
    grid=both,
    ymajorgrids=true,
    xmajorgrids=true,
    grid style=dashed,
    mark options=solid,
    width=1\columnwidth, height=7cm,
    thick,
        xmin=2,
        ymax=2,
    ymin=10e-12,
    mark size=3,
    legend style={
      anchor={center},
      cells={anchor=west},
      mark options=solid,
      column sep= 2mm,
      font=\fontsize{7pt}{7.2}\selectfont,
    },
    legend to name=ECP-SC,
    legend columns=2,
]

\addplot[
    color=black,
    dashed,
    mark=o,
    thick,
    mark size=3,
]
table {
2 0.498911
2.5 0.498189
3 0.496822
3.5 0.490951
4 0.368286
4.2 0.0911823
4.4 0.0108268
4.6 0.000443529
4.8 2.47482e-005
5 2.59482e-007
5.2 5.53e-011
};
\addlegendentry{Proposed SC-HD $N=512^2$.}

\addplot[
    color=black,
    mark=o,
    thick,
    mark size=3,
]
table {
3 0.495312
3.5 0.466311
3.75 0.262027
4 0.023337
4.5 1.15285e-005
4.75 2.0349e-010
};
\addlegendentry{Proposed SCL-HD $N=512^2$.}

%
%
%\addplot[
%    color=magenta,
%    dashed,
%    mark=o,
%    thick,
%    mark size=3,
%]
%table {
%3 0.319241
%4 0.113033
%5 0.0130446
%5.5 0.00247704
%6 0.000544218
%6.2 0.000263274
%6.4 9.74377e-005
%6.6 3.62341e-005
%6.8 2.29938e-005
%};
%\addlegendentry{BER $N=1024$, P-SC.}

%\addplot[
%    color=black,
%    mark=x,
%    thick,
%    mark size=3,
%]
%table {
%2 1
%2.5 1
%3 1
%3.5 1
%4 1
%4.2 1
%4.4 0.757
%4.6 0.081
%4.8 0.00405515
%5 9.00025e-005
%5.2 2.11e-006
%};
%\addlegendentry{BLER $N=262144$, P-SC.}

%\addplot[
%    color=magenta,
%    mark=x,
%    thick,
%    mark size=3,
%]
%table {
%3 0.995
%4 0.735
%5 0.157
%5.5 0.039
%6 0.00952381
%6.2 0.004329
%6.4 0.00164636
%6.6 0.000637511
%6.8 0.000332358
%};
%\addlegendentry{BLER $N=1024$, P-SC.}

\addplot[
    color=red,
    dashed,
    mark=o,
    thick,
    mark size=3,
]
table {
2 0.395054
2.5 0.226718
3 0.0630855 
3.5 0.00644643 
4 0.000693132
4.5 2.40005e-005
5 7.26267e-007
5.5 6.41087e-008
};
\addlegendentry{Standard SC $N=1024$.}

\addplot[
    color=red,
    mark=o,
    thick,
    mark size=3,
]
table {
2 0.238005
2.5 0.0507105
3 0.00258064
3.5 0.000111466
4 1.67805e-005
4.5 2.77272e-006
5 6.79208e-007
};
\addlegendentry{Standard SCL $N=1024$.}

\addplot[
    color=blue,
    dashed,
    mark=o,
    thick,
    mark size=3,
]
table {
2 0.450206
2.5 0.293864
3 0.0752908
3.5 0.00421209
4 6.17775e-005
4.5 9.94814e-007
5   3.73047e-008
5.5 1.08778e-009
};
\addlegendentry{Standard SC $N=2048$.}

\addplot[
    color=blue,
    mark=o,
    thick,
    mark size=3,
]
table {
2 0.342565
2.5 0.0707793
3 0.0021192
3.5 1.56478e-005
4 5.58646e-007
4.5 3.20751e-008
5.0 7.55389e-009
};
\addlegendentry{Standard SCL $N=2048$.}

\addplot[color=blue,
	thick,
		forget plot]
table {
4.91 10
4.91 0.0000000000000000000000000001
};

\addplot[color=blue,
	dashed,
	thick,
	forget plot]
table {
5.02 10
5.02 0.00000000000000000000000000001
};

\addplot[color=red,
	thick,
		forget plot]
table {
4.99 10
4.99 0.0000000000000000000000000001
};

\addplot[color=red,
	dashed,
	thick,
	forget plot]
table {
5.05 10
5.05 0.00000000000000000000000000001
};

\end{axis}
\end{tikzpicture}}
   \ref{ECP-SC}
  \\
  \vspace{2pt}
  \caption{BER comparison between standard polar decoding (SC and SCL) and proposed two-step decoding (SC and SCL, HD). Codes of rate $R=(7/8)^2$. $L=8$, $t=4$. Vertical lines represent the $E_b/N_0$ from which the proposed decoder is faster than the standard SC-based one. }
  \label{fig:ECP-SC}
\end{figure}

%\begin{figure}
%  \centering
%  \includegraphics[width=0.47\textwidth]{figures/ECP1.eps}
%  \caption{BER for SC and P-SC-HD with comparison of hard decisions, for codes of rate $R=(7/8)^2$. \fixme{Change label to P-SC-HD}{}}
%  \label{fig:ECP-SC}
%\end{figure}
%

Figure~\ref{fig:ECP-SC} portrays the bit error rate (BER) for code $N = 512^2$ with rate $R = (7/8)^2$ under the proposed SC-HD and SCL-HD decoding, with parameters $t=4$ and $L=8$, and no CRC. 
%The selected frozen set is optimal for product decoding. 
The frozen set is selected according to the procedure presented in Section~\ref{sec:PPC2PC}. 
As a reference, Figure~\ref{fig:ECP-SC} displays also curves obtained with SC and SCL decoding of polar codes of length $N=1024$ and $N=2048$, with the same rate $R=(7/8)^2$, designed according to \cite{arikan}.
The longer code imposes a steeper slope with respect to standard polar decoding in both SC-HD and SCL-HD: the BER curves are shown to cross at around $\text{BER} \simeq 10^{-7}$. 
Comparison with such different code lengths is made possible by the fact that the speedup achieved by SC-HD and SCL-HD over standard SC and SCL allows to decode longer codes within the same time constraint. 
The nature of this comparison is linked to parameters $t_{avg}$ and $\gamma$ introduced in Section \ref{subsec:latency}, that reflect the performance of the first decoding step. 
Through simulation, we have observed that the average number of iterations $t_{avg}$ tends to $1$ and $\gamma$ tends to $0$ as $E_b/N_0$ increases.
The slope with which $\gamma$ tends to $0$ changes depending on the value of $t$; as $t$ increases, so does the steepness of the $\gamma$ curve. 
Moreover, the slope and waterfall region for both parameters is deeply affected by the choice of the frozen set. 
As the effectiveness of the first decoding step increases, the average number of iterations and the number of times the second decoding step is activated decreases, leading to lower decoding latency.  

\begin{figure}
  \centering
    \begin{tikzpicture}
  \pgfplotsset{
    label style = {font=\fontsize{9pt}{7.2}\selectfont},
    tick label style = {font=\fontsize{7pt}{7.2}\selectfont}
  }

\begin{axis}[
	scale = 1,
    ymode=log,
    xlabel={$E_b/N_0$ [\text{dB}]}, xlabel style={yshift=0.4em},
    ylabel={$\gamma$}, ylabel style={yshift=-0.75em},
    grid=both,
    ymajorgrids=true,
    xmajorgrids=true,
    grid style=dashed,
    mark options=solid,
    width=1\columnwidth, height=7cm,
    thick,
        xmin=3,
    mark size=3,
    legend style={
      anchor={center},
      cells={anchor=west},
      mark options=solid,
      column sep= 2mm,
      font=\fontsize{7pt}{7.2}\selectfont,
    },
    legend to name=Gamma-SCL-SD,
    legend columns=2,
]

\addplot[
    color=black,
    mark=o,
    thick,
    mark size=3,
]
table {
1 1
1.5 1
2 1
2.5 1
3 0.9988
3.5 0.9836
4 0.8852
4.5 0.6546
5 0.3677
5.5 0.161719
6 0.0590114
6.5 0.0190113
};
\addlegendentry{$N=32^2$, HD, $L=8$}

\addplot[
    color=black,
    dashed,
    mark=o,
    thick,
    mark size=3,
]
table {
1 1
1.5 1
2 0.9988
2.5 0.9708
3 0.8303
3.5 0.547
4 0.2621
4.5 0.1012
5 0.0297
5.5 0.00771052
6 0.00190521
6.5 0.000284529
};
\addlegendentry{$N=32^2$, SD, $L=8$}

\addplot[
    color=red,
    mark=o,
    thick,
    mark size=3,
]
table {
1 1
1.5 1
2 1
2.5 1
3 1
3.5 1
4 0.967
4.5 0.222118
5 0.00823158
};
\addlegendentry{$N=128^2$, HD, $L=8$}

\addplot[
    color=red,
    dashed,
    mark=o,
    thick,
    mark size=3,
]
table {
1 1
1.5 1
2 1
2.5 1
3 1
3.5 1
4 0.888
4.5 0.277
5 0.0476993
};
\addlegendentry{$N=128^2$, SD, $L=8$}

\addplot[
    color=blue,
    mark=x,
    thick,
    mark size=3,
]
table {
1 1
1.5 1
2 1
2.5 1
3 1
3.5 1
4 0.969
4.5 0.224814
5 0.00615657
};
\addlegendentry{$N=128^2$, HD, $L=32$}

\addplot[
    color=blue,
    dashed,
    mark=x,
    thick,
    mark size=3,
]
table {
1 1
1.5 1
2 1
2.5 1
3 1
3.5 0.991
4 0.625
4.5 0.146425
5 0.0010215
};
\addlegendentry{$N=128^2$, SD, $L=32$}

%\addplot[
%    color=red,
%    thick,
%]
%table {
%2 0.238005
%2.5 0.0507105
%3 0.00258064
%3.5 0.000111466
%4 1.67805e-005
%4.5 2.77272e-006
%5 6.79208e-007
%};
%\addlegendentry{BER $N=1024$, SCL, $L=8$.}
%
%
%1 0.463254
%1.5 0.392485
%2 0.171691
%2.5 0.0225791
%3 0.00102482
%3.5 0.000110716
%4 1.91923e-005
%4.5 3.14593e-006

\end{axis}
\end{tikzpicture}
    \\
    \ref{Gamma-SCL-SD}
  \caption{Evolution of $\gamma$ with SCL-HD and SCL-SD for different code lengths and list sizes, $R_r=R_c=7/8$, $t=4$.}
  \label{fig:Gamma-SC-SD}
\end{figure}
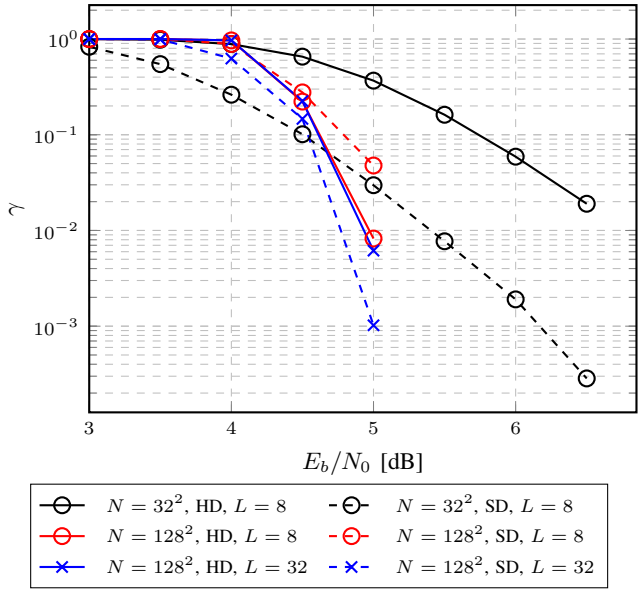

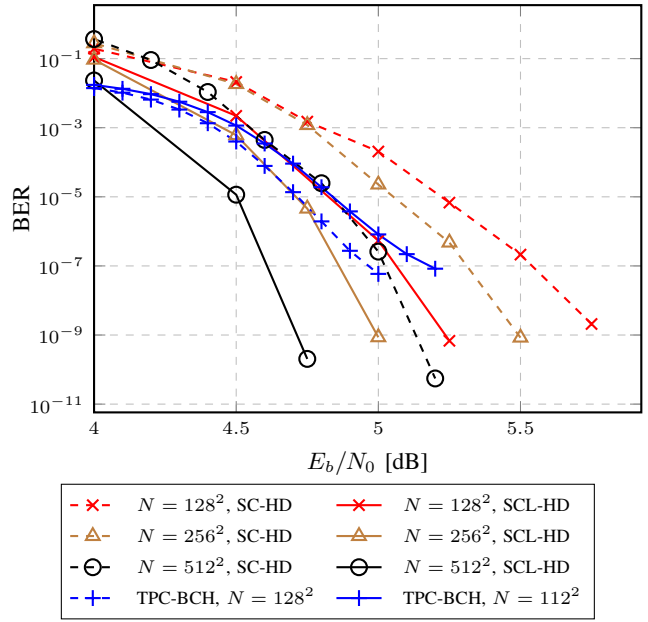
\begin{figure}
  \centering
   \scalebox{1}{\begin{tikzpicture}
  \pgfplotsset{
    label style = {font=\fontsize{9pt}{7.2}\selectfont},
    tick label style = {font=\fontsize{7pt}{7.2}\selectfont}
  }

\begin{axis}[
	scale = 1,
    ymode=log,
    xlabel={$E_b/N_0$ [\text{dB}]}, xlabel style={yshift=0.4em},
    ylabel={BER}, ylabel style={yshift=-0.75em},
    grid=both,
    ymajorgrids=true,
    xmajorgrids=true,
    grid style=dashed,
    mark options=solid,
    width=1\columnwidth, height=7cm,
    thick,
        xmin=4,
    mark size=3,
    legend style={
      anchor={center},
      cells={anchor=west},
      mark options=solid,
      column sep= 2mm,
      font=\fontsize{7pt}{7.2}\selectfont,
    },
    legend to name=ECP-BCH,
    legend columns=2,
]

\addplot[
    color=red,
    dashed,
    mark=x,
    thick,
    mark size=3,
]
table {
2 0.490758
2.5 0.483431
3 0.460188
3.5 0.388332
4 0.194146
4.5 0.0211321
4.75 0.00150287
5 0.000205055 
5.25 6.80267e-006
5.5 2.12518e-007
5.75 2.08436e-009
};
%\addlegendentry{$N=128^2$, P-SC-HD}
\addlegendentry{$N=128^2$, SC-HD}

\addplot[
    color=red,
    mark=x,
    thick,
    mark size=3,
]
table {
1 0.49513
1.5 0.492959
2 0.489671
2.5 0.478765
3 0.444164
3.5 0.336736
4 0.111662
4.5 0.0022104
5 5.39873e-007
5.25 6.80581e-010
};
%\addlegendentry{$N=128^2$, P-SCL-HD, $L=8$}
\addlegendentry{$N=128^2$, SCL-HD}

\addplot[
    color=brown,
    dashed,
    mark=triangle,
    thick,
    mark size=3,
]
table {
2 0.496861
2.5 0.494545
3 0.489053
3.5 0.460976
4 0.27815
4.5 0.0183717
4.75 0.00119348
5 2.27109e-005
5.25 4.73808e-007
5.5 8.41302e-010
};
%\addlegendentry{$N=256^2$, P-SC-HD}
\addlegendentry{$N=256^2$, SC-HD}

\addplot[
    color=brown,
    mark=triangle,
    thick,
    mark size=3,
]
table {
2 0.49682
2.5 0.493511
3 0.451022
3.5 0.409553
4 0.0933489
4.5 0.000608099
4.75 4.56002e-006
5 8.61443e-010
};
%\addlegendentry{$N=256^2$, P-SC-HD}
\addlegendentry{$N=256^2$, SCL-HD}

\addplot[
    color=black,
    dashed,
    mark=o,
    thick,
    mark size=3,
]
table {
2 0.498911
2.5 0.498189
3 0.496822
3.5 0.490951
4 0.368286
4.2 0.0911823
4.4 0.0108268
4.6 0.000443529
4.8 2.47482e-005
5 2.59482e-007
5.2 5.53e-011
};
%\addlegendentry{$N=512^2$, P-SC-HD}
\addlegendentry{$N=512^2$, SC-HD}

\addplot[
    color=black,
    mark=o,
    thick,
    mark size=3,
]
table {
3 0.495312
3.5 0.466311
3.75 0.262027
4 0.023337
4.5 1.15285e-005
4.75 2.0349e-010
};
%\addlegendentry{$N=512^2$, P-SCL-HD, $L=8$}
\addlegendentry{$N=512^2$, SCL-HD}

%

%\addplot[
%    color=red,
%    mark=triangle,
%    thick,
%    mark size=3,
%]
%table {
%1 0.498023
%1.5 0.497401
%2 0.495772
%2.5 0.489351
%3 0.465263
%3.5 0.384472
%4 0.152706
%4.5 0.00448916
%5 5.21677e-007
%};
%\addlegendentry{$N=128^2$, P-SCL-SD, $L=8$}

%\addplot[
%    color=blue,
%    mark=+,
%    thick,
%    mark size=3,
%]
%table {
%4	0.0236642947384067
%4.1 0.0219650713436385
%4.2 0.0202040167954816
%4.3 0.0183915539536266
%4.4 0.0164932372175981
%4.5 0.0145160337395957
%4.6 0.0124732647146254
%4.7 0.0104151679548157
%4.8 0.00828723246135553
%4.9 0.00597196417954816
%5.0 0.00354509979071124
%5.1 0.00143962787255211
%5.2 0.000302904755984451
%5.3 3.31305419818430e-05
%5.4 1.85795715475799e-06
%5.5 8.67757253505094e-08
%};
%\addlegendentry{TPC-BCH, $N=248^2$, HD}
%
%\addplot[
%    color=blue,
%    mark=+,
%    thick,
%    mark size=3,
%]
%table {
%3 0.0306346425386445	
%3.1 0.0285118534482759	
%3.2 0.0261857164090369	
%3.3 0.0235438280321046	
%3.4 0.0204300683709869	
%3.5 0.0163631465517241	
%3.6 0.0103123142092747	
%3.7 0.00313498184336472	
%3.8 0.000246669232898126	
%3.9 4.83764363759368e-06
%4 6.41815778317579e-08
%};
%\addlegendentry{TPC-BCH, $N=248^2$, SD}

\addplot[
    color=blue,
    dashed,
    mark=+,
    thick,
    mark size=3,
]
table {
4	0.0140284598214286	
4.1 0.0103608447327598
4.2 0.00651895980354759	
4.3 0.00335653663938103	
4.4 0.00134928924997719	
4.5 0.000398927343790383	
4.6 7.84439598366926e-05	
4.7 1.36483987331019e-05	
4.8 1.94636379111162e-06
4.9 2.73953480141565e-07	
5.0 5.88032604138671e-08
};
\addlegendentry{TPC-BCH, $N=128^2$}

\addplot[
    color=blue,
    mark=+,
    thick,
    mark size=3,
]
table {
4 0.0173934769936436
4.1 	0.0133715241923360
4.2 	0.00939168800372175	
4.3 0.00565532497546197
4.4 	0.00282747955397372	
4.5 0.00115558367368054	
4.6 0.000349706844515965
4.7 	9.15251892351037e-05	
4.8 1.89580882440941e-05	
4.9 3.78876085333684e-06	
5 8.16998021171152e-07	
5.1 2.19298590536281e-07	
5.2 8.30785075242692e-08
};
\addlegendentry{TPC-BCH, $N=112^2$}

\end{axis}
\end{tikzpicture}}
   \ref{ECP-BCH}
  \\
  \vspace{2pt}
  \caption{BER comparison between BCH TPC and polar product codes, $R=(7/8)^2$, $t=4$. For SCL, $L=8$.}
  \label{fig:BER-BCH}
\end{figure}

Whereas the impact of $t_{avg}$ on the decoding latency is minimal, since usually $\delta_{N_r}<<\delta_N$, the value of $\gamma$ plays a major role on the decoding speed. 
%It is thus important to identify the conditions under which the proposed decoding approach is faster than standard polar decoding.
Let us consider the case of SC decoding of a code of length $N$, for which $\delta^{\rm SC}_N=2N-2$. 
According to (\ref{eq:timeHD}), SC-HD decoding of a code of length $N_r^2$ is faster if
\begin{equation*}
t_{avg}(2N_r-2) + \gamma (2N_r^2 -2) < 2N-2~,
\end{equation*}
that becomes after basic transformations
\begin{equation}\label{eq:SCgamma}
\gamma < \frac{t_{avg}(1-N_r) + N-1}{N_r^2-1} \approx \frac{N/N_r - t_{avg}}{N_r}~.
\end{equation}
Following the same reasoning in case of SCL, where $\delta^{\rm SCL}_N=N(2+R)-2$, decoding the length-$N$ code through the proposed two-step decoder is faster than common polar decoding if 
\begin{equation} \label{eq:SCLgamma}
\gamma \lessapprox \frac{N/N_r(2+R) - t_{avg}(2+R_r)}{N_r(2+R_r^2)}~.
\end{equation}
In Figure \ref{fig:ECP-SC}, the code with $N_r=512$ is compared to those with $N=1024$ and $N=2048$, for both SC and SCL-based decoding. 
Four vertical lines are shown, indicating the $E_b/N_0$ from which the proposed decoding is faster than standard SC and SCL decoding of codes of length $N=1024$ and $N=2048$, according to (\ref{eq:SCgamma}) and (\ref{eq:SCLgamma}).
It can be seen that at these $E_b/N_0$ points, the proposed decoding approach has substantially lower BER than its competitor. 
An exception is found for SC decoding with $N=2048$, for which the BER matches that of SC-HD of $N_r=512$. 
The steeper slope of the latter guarantees a significant advantage over the former at all higher $E_b/N_0$.

%
%\begin{figure}
%  \centering
%  \includegraphics[width=0.47\textwidth]{figures/ECP2.eps}
%  \caption{BER comparison for SCL and P-SCL-HD, for codes of rate $R=(7/8)^2$, $L=8$, $t=4$. \fixme{Change label to P-SCL-HD}{}}
%  \label{fig:ECP-SCL}
%\end{figure}
%

%\begin{figure}
%  \centering
%   \scalebox{1}{\input{figures/ECP-SD1.tikz}}
%   \ref{ECP-SD-SCL1}
%  \\
%  \vspace{2pt}
%  \caption{BER comparison between P-SCL-HD and P-SCL-SD for different code lengths and list sizes, $R_r=R_c=7/8$, $t=4$}
%  \label{fig:BER-SCL-SD}
%\end{figure}

If we consider the proposed SCL-HD and SCL-SD approaches, simulation results have shown approximately the same error-correction performance. 
This is because the second decoding step is the same regardless of the information exchange criterion within the first step. 
Consequently, notwithstanding how well the first step is able to decode errors, if residual errors are detected the second step is activated. 
However, SCL-SD can bring substantial reduction in the number of times in which the second step is needed, thus proving its improved effectiveness with respect to SCL-HD.
The evolution of $\gamma$ with both SCL-HD and SCL-SD, for different codes and list sizes, is detailed in Figure \ref{fig:Gamma-SC-SD}. 
For a code length of $N_r=N_c=32$, with $L=8$, a large difference can be observed between the performance of the two information exchange techniques: the soft information exchange in SCL-SD is able to greatly improve the effectiveness of the first decoding step. 
However, for the larger $N_r=N_c=128$ code, SCL-SD with $L=8$ performs worse than the SCL-HD version. 
This is due to the fact that the simple soft information calculation in (\ref{eq:soft}) is based on the difference between $\PM$s, which assume up to $L$ different values: consequently, the computed $\Lambda$s can assume a very limited set of values within the same codeword. 
This limitation degrades the error-correction performance of SCL-SD when the ratio between $N_r$ ($N_c$) and $L$ is too large, since it prevents $\Lambda$ to distinguish between more and less reliable bits. 
With $L=32$ and $N_r=N_c=128$, SCL-SD has a lower activation rate of the second decoding step than SCL-HD.

Figure \ref{fig:BER-BCH} plots the BER for product polar codes of rate $R=(7/8)^2$, decoded with SC-HD and SCL-HD, and that of two turbo product codes (TPCs) with the same rate with Bose-Chaudhuri-Hocquenghem (BCH) codes \cite{Bose_IC60} as component codes. 
In particular, the $N=128^2$ TPC is based on a double-error-correcting BCH shortened from the BCH of length $256$. 
The $N=112^2$ TPC is instead constructed by shortening the double-error-correcting BCH code of length $128$. 
BCH component codes have been decoded through bounded distance decoding, with hard decision iterations.
Product polar codes of the same length show a waterfall region at higher $E_b/N_0$ than TPCs, but with a steeper slope. 
Polar codes, moreover, do not show an error floor \cite{pol_err_floor}, that is instead encountered in TPCs. 
The high flexibility with which the rate of polar codes can be changed is also an advantage over polynomial codes.

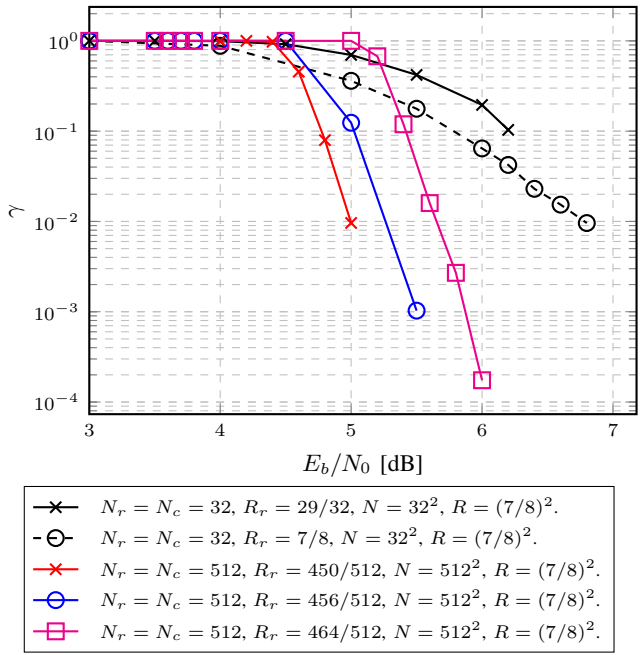
\begin{figure}
  \centering
    \begin{tikzpicture}
  \pgfplotsset{
    label style = {font=\fontsize{9pt}{7.2}\selectfont},
    tick label style = {font=\fontsize{7pt}{7.2}\selectfont}
  }

\begin{axis}[
	scale = 1,
    ymode=log,
    xlabel={$E_b/N_0$ [\text{dB}]}, xlabel style={yshift=0.4em},
    ylabel={$\gamma$}, ylabel style={yshift=-0.75em},
    grid=both,
    ymajorgrids=true,
    xmajorgrids=true,
    grid style=dashed,
    mark options=solid,
    width=1\columnwidth, height=7cm,
    thick,
        xmin=3,
    mark size=3,
    legend style={
      anchor={center},
      cells={anchor=west},
      mark options=solid,
      column sep= 2mm,
      font=\fontsize{7pt}{7.2}\selectfont,
    },
    legend to name=Gamma-SC,
    legend columns=1,
]

\addplot[
    color=black,
    mark=x,
    thick,
    mark size=3,
]
table {
3.0 1
3.5 1
4.0 0.993
4.5 0.91846
5.0 0.698785
5.5 0.419809
6.0 0.19512
6.2 0.103323
};
\addlegendentry{$N_r=N_c=32$, $R_r=29/32$, $N=32^2$, $R=(7/8)^2$.}

\addplot[
    color=black,
    dashed,
    mark=o,
    thick,
    mark size=3,
]
table {
3 1
4 0.882
5 0.361
5.5 0.177
6 0.064381
6.2 0.0422511
6.4 0.0229832
6.6 0.0153513
6.8 0.00959851
};
\addlegendentry{$N_r=N_c=32$, $R_r=7/8$, $N=32^2$, $R=(7/8)^2$.}

\addplot[
    color=red,
    mark=x,
    thick,
    mark size=3,
]
table {
4.0 1
4.2 1
4.4 0.976243
4.6 0.4555399
4.8 0.0794931
5.0 0.0096238
};
\addlegendentry{$N_r=N_c=512$, $R_r=450/512$, $N=512^2$, $R=(7/8)^2$.}

\addplot[
    color=blue,
    mark=o,
    thick,
    mark size=3,
]
table {
3 1
3.5 1
3.6 1
3.7 1
3.8 1
4.0 1
4.5 1
5.0 0.12393
5.5 0.0010281
};
\addlegendentry{$N_r=N_c=512$, $R_r=456/512$, $N=512^2$, $R=(7/8)^2$.}

\addplot[
    color=magenta,
    mark=square,
    thick,
    mark size=3,
]
table {
3 1
3.5 1
3.6 1
3.7 1
3.8 1
4.0 1
4.5 1
5.0 1
5.2 0.67529
5.4 0.119188
5.6 0.0159768
5.8 0.00269635
6.0 0.00017433
};
\addlegendentry{$N_r=N_c=512$, $R_r=464/512$, $N=512^2$, $R=(7/8)^2$.}

\end{axis}
\end{tikzpicture}
    \\
    \ref{Gamma-SC}
  \caption{Evolution of $\gamma$ for different code lengths and rates, SC-HD decoding, $t=4$, $N_r=N_c$, $R_r=R_c$, and mixed frozen set design.}
  \label{fig:Gamma-SC-mixed}
\end{figure}

\begin{figure}
  \centering
   \scalebox{1}{\begin{tikzpicture}
  \pgfplotsset{
    label style = {font=\fontsize{9pt}{7.2}\selectfont},
    tick label style = {font=\fontsize{7pt}{7.2}\selectfont}
  }

\begin{axis}[
	scale = 1,
    ymode=log,
    xlabel={$E_b/N_0$ [\text{dB}]}, xlabel style={yshift=0.4em},
    ylabel={BER}, ylabel style={yshift=-0.75em},
    grid=both,
    ymajorgrids=true,
    xmajorgrids=true,
    grid style=dashed,
    mark options=solid,
    width=1\columnwidth, height=7cm,
    thick,
        xmin=3,
    mark size=3,
    legend style={
      anchor={center},
      cells={anchor=west},
      mark options=solid,
      column sep= 2mm,
      font=\fontsize{7pt}{7.2}\selectfont,
    },
    legend to name=ECP-mixed,
    legend columns=1,
]

\addplot[
    color=black,
    mark=x,
    thick,
    mark size=3,
]
table {
3 0.27208
3.5 0.0996633
4 0.0215893
4.5 0.001861
5 0.000178182
5.5 6.21686e-006
%6 5.12699e-007
%6.2 2.76721e-007
};
\addlegendentry{$N_r=N_c=32$, $R_r=29/32$, $N=32^2$, $R=(7/8)^2$.}

\addplot[
    color=black,
    dashed,
    mark=o,
    thick,
    mark size=3,
]
table {
3 0.255099
3.5 0.179982
4 0.0936964
4.5 0.0298444
5 0.00800383
5.5 0.000678694
%6 0.000233515
%6.2 0.000116861
};
\addlegendentry{$N_r=N_c=32$, $R_r=7/8$, $N=32^2$, $R=(7/8)^2$.}

\addplot[
    color=red,
    mark=x,
    thick,
    mark size=3,
]
table {
3 0.497414
3.5 0.493105
4 0.166454
4.2 0.00145455
4.4 1.01317e-005
};
\addlegendentry{$N_r=N_c=512$, $R_r=450/512$, $N=512^2$, $R=(7/8)^2$.}

\addplot[
    color=blue,
    mark=o,
    thick,
    mark size=3,
]
table {
3 0.497322
3.5 0.494284
3.6 0.475776
3.7 0.221169
3.8 0.0335725
4.0 0.000134868
};
\addlegendentry{$N_r=N_c=512$, $R_r=456/512$, $N=512^2$, $R=(7/8)^2$.}

\addplot[
    color=magenta,
    mark=square,
    thick,
    mark size=3,
]
table {
3 0.497497
3.5 0.420388
3.6 0.125802
3.7 0.0158346
3.8 0.000524034
};
\addlegendentry{$N_r=N_c=512$, $R_r=464/512$, $N=512^2$, $R=(7/8)^2$.}

\end{axis}
\end{tikzpicture}}
   \ref{ECP-mixed}
  \\
  \vspace{2pt}
  \caption{BER for different code lengths and rates, SC-HD decoding, $t=4$, $N_r=N_c$, $R_r=R_c$, and mixed frozen set design.}
  \label{fig:BER-SC-mixed}
\end{figure}
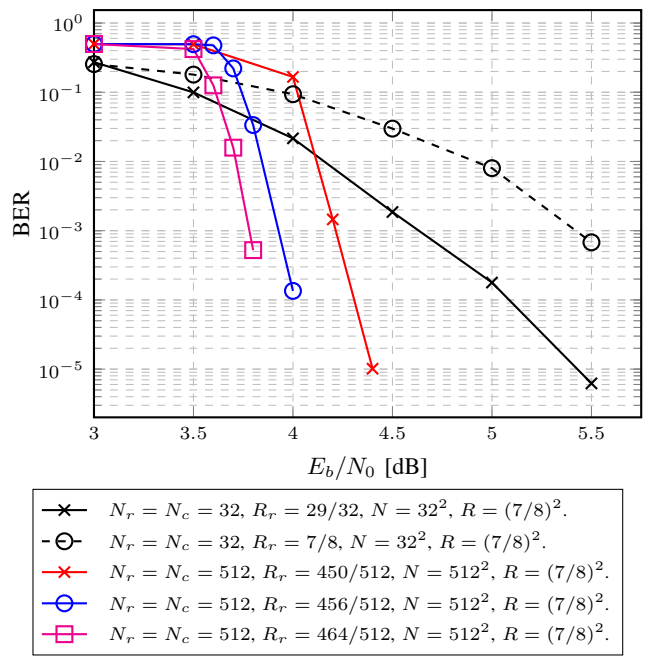

The mixed frozen set design technique proposed in Section \ref{sec:frozenSetDesign} has been evaluated through extensive simulations as well.
Figure \ref{fig:Gamma-SC-mixed} shows the evolution of $\gamma$ with different initial rates $R_r$ and $R_c$, for SC-HD decoding. 
It can be observed that as the difference between $R_r\cdot R_c$ and $R$ increases, $\gamma$ increases as well. 
A higher $R_r$ and $R_c$ lead to less powerful product code decoding, and thus a higher fraction $\gamma$ of instances in which Step 2 is needed. 
As foreseen, from Fig. \ref{fig:BER-SC-mixed}, we can see that the BER improves as more frozen bits are selected to be optimal for polar decoding. 
Similar observations are made for SCL-HD and SCL-SD decoding methods.

\section{Conclusions}\label{sec:conc}

In this paper, we highlighted the dual nature of polar codes as a particular case of product codes. 
According to this interpretation, the product of two polar codes results again in a polar code, and a polar code can be seen as an irregular product code. 
This allowed us to propose a novel two-step decoder for product polar codes heavily reducing the decoding latency for very long codes. 
We proposed a hard decision decoder based on this technique, along with a soft decision decoder based on the nature of the path metrics calculated during SCL decoding. 
Moreover, we proposed a frozen set design that exploits the dual nature of the resulting product polar code to trade-off between decoding performance and latency. 
Performance analysis and simulations show that the high throughput of the proposed decoding approach enables the targeting of very long codes, while granting good error correction performance suitable for optical communications.

%\bibliographystyle{IEEEbib}
%\bibliography{IEEEabrv,refs}

\begin{thebibliography}{10}

\bibitem{arikan}
E.~Ar{\i}kan,
\newblock ``Channel polarization: A method for constructing capacity-achieving
  codes for symmetric binary-input memoryless channels,''
\newblock {\em IEEE Transactions on Information Theory}, vol. 55, no. 7, pp.
  3051--3073, July 2009.

\bibitem{tal_list}
I.~Tal and A.~Vardy,
\newblock ``List decoding of polar codes,''
\newblock {\em IEEE Transactions on Information Theory}, vol. 61, no. 5, pp.
  2213--2226, May 2015.

\bibitem{elias}
P.~Elias,
\newblock ``Error-free coding,''
\newblock {\em Transactions of the IRE Professional Group on Information
  Theory}, vol. 4, no. 4, pp. 29--37, 1954.

\bibitem{pc_inner}
M.~Seidl and J.~B. Huber,
\newblock ``Improving successive cancellation decoding of polar codes by usage
  of inner block codes,''
\newblock in {\em IEEE International Symposium on Turbo Codes and Iterative
  Information Processing (ISTC)}, Brest, France, September 2010.

\bibitem{BP_pc}
J.~Guo, M.~Qin, A.~G. I~Fabregas, and P.~H. Siegel,
\newblock ``Enhanced belief propagation decoding of polar codes through
  concatenation,''
\newblock in {\em IEEE International Symposium on Information Theory (ISIT),
  2014}, Honolulu, HI, USA, June 2014.

\bibitem{par_conc_sys_pol}
D.~Wu, A.~Liu, Y.~Zhang, and Q.~Zhang,
\newblock ``Parallel concatenated systematic polar codes,''
\newblock in {\em Electronics Letters}, 2015, vol.~52, pp. 43--45.

\bibitem{PRODPOL_FIRST}
V.~Bioglio, C.~Condo, and I.~Land,
\newblock ``Construction and decoding of product codes with non-systematic
  polar codes,''
\newblock in {\em IEEE Wireless Communications and Networking Conference
  (WCNC)}, Marrakech, Morocco, April 2019.

\bibitem{Arikan_2D}
E.~Ar{\i}kan and G.~Markarian,
\newblock ``Two-dimensional polar coding,''
\newblock in {\em Int. Symp. on Commun. Theory and App. (ISCTA)}, July 2009,
  pp. 1--3.

\bibitem{Product2014}
H.~{Mahdavifar}, M.~{El-Khamy}, J.~{Lee}, and I.~{Kang},
\newblock ``Fast multi-dimensional polar encoding and decoding,''
\newblock in {\em 2014 Information Theory and Applications Workshop (ITA)}, Feb
  2014, pp. 1--5.

\bibitem{Trifonov_1}
P.~{Trifonov},
\newblock ``Efficient design and decoding of polar codes,''
\newblock {\em IEEE Transactions on Communications}, vol. 60, no. 11, pp.
  3221--3227, November 2012.

\bibitem{Trifonov_3}
P.~{Trifonov} and V.~{Miloslavskaya},
\newblock ``Polar subcodes,''
\newblock {\em IEEE Journal on Selected Areas in Communications}, vol. 34, no.
  2, pp. 254--266, Feb 2016.

\bibitem{SCAN_pc}
U.~U. Fayyaz and J.~R. Barry,
\newblock ``Low-complexity soft-output decoding of polar codes,''
\newblock {\em IEEE Journal on Selected Areas in Communications}, vol. 32, no.
  5, pp. 958--966, 2014.

\bibitem{par_sys_list}
Z.~Liu, K.~Niu, and J.~Lin,
\newblock ``Parallel concatenated systematic polar code based on soft
  successive cancellation list decoding,''
\newblock in {\em IEEE International Symposium on Wireless Personal Multimedia
  Communications (WPMC)}, Yogyakarta, Indonesia, December 2017.

\bibitem{KoikeAkinoIrregularPT}
T.~Koike-Akino, C.~Cao, Y.~Wang, K.~Kojima, D.~S. Millar, and K.~Parsons,
\newblock ``Irregular polar turbo product coding for high-throughput optical
  interface,''
\newblock in {\em Optical Fiber Communication Conference and Exhibition (OFC)},
  San Diego, CA, USA, 2018, p. March.

\bibitem{polar_const}
H.~Vangala, E.~Viterbo, and Y.~Hong,
\newblock ``A comparative study of polar code constructions for the {AWGN}
  channel,''
\newblock in {\em arXiv preprint arXiv:1501.02473}, 2015.

\bibitem{mcw_sloane}
F.~J. MacWilliams and N.~J.~A. Sloane,
\newblock {\em The theory of error-correcting codes},
\newblock Elsevier, 1977.

\bibitem{block_turbo}
R.~M. Pyndiah,
\newblock ``Near-optimum decoding of product codes: Block turbo codes,''
\newblock {\em IEEE Transactions on communications}, vol. 46, no. 8, pp.
  1003--1010, 1998.

\bibitem{irr_prod}
M.~Alipour, O.~Etesami, G.~Maatouk, and A.~Shokrollahi,
\newblock ``Irregular product codes,''
\newblock in {\em IEEE Information Theory Workshop (ITW)}, Lausanne,
  Switzerland, Sept. 2012.

\bibitem{CHASE}
D.~Chase,
\newblock ``Class of algorithms for decoding block codes with channel
  measurement information,''
\newblock {\em IEEE Transactions on Information Theory}, vol. 18, no. 1, pp.
  170--182, January 1972.

\bibitem{balatsoukas_SCL_HW}
A.~Balatsoukas-Stimming, A.~J. Raymond, W.~J. Gross, and A.~Burg,
\newblock ``Hardware architecture for list successive cancellation decoding of
  polar codes,''
\newblock {\em IEEE Transactions on Circuits and Systems II: Express Briefs},
  vol. 61, no. 8, pp. 609--613, August 2014.

\bibitem{hashemi_FSSCL_TSP}
S.~A. Hashemi, C.~Condo, and W.~J. Gross,
\newblock ``Fast and flexible successive-cancellation list decoders for polar
  codes,''
\newblock {\em IEEE Transactions on Signal Processing}, vol. 65, no. 21, pp.
  5756--5769, October 2017.

\bibitem{sarkis}
G.~Sarkis, P.~Giard, A.~Vardy, C.~Thibeault, and W.J. Gross,
\newblock ``Fast polar decoders: Algorithm and implementation,''
\newblock {\em IEEE Journal on Selected Areas in Communications}, vol. 32, no.
  5, pp. 946--957, May 2014.

\bibitem{hashemi_SSCL}
S.~A. Hashemi, C.~Condo, and W.~J. Gross,
\newblock ``Simplified successive-cancellation list decoding of polar codes,''
\newblock in {\em IEEE International Symposium on Information Theory (ISIT)},
  Barcelona, Spain, July 2016.

\bibitem{Condo_GLOBECOM}
C.~Condo, V.~Bioglio, and I.~Land,
\newblock ``Generalized fast decoding of polar codes,''
\newblock in {\em IEEE Global Communications Conference (GLOBECOM)}, Abu Dhabi,
  UAE, Dec. 2018.

\bibitem{Bose_IC60}
R.C. Bose and D.K. Ray-Chaudhuri,
\newblock ``On a class of error correcting binary group codes,''
\newblock {\em Information and Control}, vol. 3, no. 1, pp. 68 -- 79, 1960.

\bibitem{pol_err_floor}
M.~Mondelli, S.~H. Hassani, and R.~L. Urbanke,
\newblock ``Unified scaling of polar codes: Error exponent, scaling exponent,
  moderate deviations, and error floors,''
\newblock {\em IEEE Transactions on Information Theory}, vol. 62, no. 12, pp.
  6698--6712, December 2016.

\end{thebibliography}

\appendices 
\section{Proof of Proposition \ref{prop:frozen}} \label{proof:PPC}
%Input and codeword matrices $U$ and $X$ have to be vectorized into row vectors $u$ and $x$ in order to show that this procedure creates a polar code. 
%Given the classical vectorization function $\vect(\cdot)$ converting matrices into column vectors, here we define an analogous linear transformation $\row(\cdot)$ converting a matrix into a row vector by juxtaposing its rows head-to-tail. 
%Equipped with this definition, we can extend a classical result of $\vect(\cdot)$ function to $\row(\cdot)$ function. 
Given the classical vectorization function $\vect(\cdot)$ converting matrices into column vectors, we begin the proof extending a classical result of $\vect(\cdot)$ function to $\row(\cdot)$ function. 
\begin{lemma}
\label{prop:row}
Given three matrices $A$, $B$, $C$, then %if $A \cdot B \cdot C$ is defined, then
\begin{equation}
\row(A \cdot B \cdot C) = \row(B) \cdot (A^T \otimes C).
\end{equation}
\begin{proof}
The compatibility of vectorization with the Kronecker product is a well known result, that is used to express matrix multiplication $A \cdot B \cdot C$ as a linear transformation $\vect(A \cdot B \cdot C) = (C^T \otimes A) \cdot \vect(B)$. 
Having $\vect(A^T) = (\row(A))^T$ by construction, then
\begin{align*}
\row(A \cdot B \cdot C) & =  (\vect((A \cdot B \cdot C)^T))^T  \\
  & =  (\vect(C^T \cdot B ^T\cdot A^T))^T  \\
  & =  ((A \otimes C^T) \cdot \vect(B^T))^T  \\
  & =  (\vect(B^T))^T \cdot  (A \otimes C^T)^T  \\
  & =  \row(B) \cdot (A^T \otimes C).
\end{align*}
\end{proof}
\end{lemma}
%Lemma~\ref{prop:row} is used to prove the main proposition of this section: 
We now define $u = \row(U)$, so that input vector $u$ has frozen bits imposed by \eqref{eq:frozen} according to the definition of input matrix $U$. 
With slight abuse of notation, we use the $\arg \min$ function to return the set of the indices of vector $z = z_c \otimes z_r$ for which the entry is zero. 
Polar codeword $x$ is calculated through Lemma~\ref{prop:row} as 
\begin{align*}
x & = \row(X) \\
  & = \row(T_{N_c}^T \cdot U \cdot T_{N_r}) \\
  & = \row(U) \cdot (T_{N_c} \otimes T_{N_r}) \\
  & = u \cdot T_N. 
\end{align*}
Finally, if $N_r = 2^{n_r}$ and $N_c = 2^{n_c}$, then $T_N = T_{N_c} \otimes T_{N_r} = T_2^{\otimes(n_c + n_r)}$, hence $T_N$ is the transformation matrix of a polar code of length $N = 2^{n_c + n_r}$.

\section{Proof of Proposition \ref{prop:frozen_small}} \label{proof:IPC}
%In the following, we call $x_r^i$ and $x_c^j$ the codewords formed by the $i$-th row and the $j$-th column of $X$ respectively, using codes $\mathcal{C}_r^i$ and $\mathcal{C}_c^j$. 
%To begin with, we provide a procedure to calculate the frozen set $\mathcal{F}_r^i$ of row code $\mathcal{C}_r^i$; in the following, we define $A^{(i)}$ as the $i$-th column of matrix $A$. 
%%We begin the analysis proving a proposition on the frozen set $\mathcal{F}_r^i$ of row code $\mathcal{C}_r^i$. 
%%with row codes with the following proposition. 
%\begin{proposition}
%\label{prop:frozen_small_r}
Given the polar code $\mathcal{C}_r^i$ defined on the $i$-th row of $X$, the bit of index $l$ belongs to the frozen set $\mathcal{F}_r^i$ only if $\mathcal{S}_l \subset \mathcal{F}$, where 
\begin{equation}
\label{eq:frozen_small_r}
\mathcal{S}_l = \arg \max (T_{N_c}^{(i)}  \otimes b_{N_r}^l)
\end{equation} 
and $b_{N_r}^l$ is the binary column vector of length $N_r$ having one in the $l$-th position and zeros elsewhere.
Given the nature of the transformation matrix, the virtual input vector $u_r^i$ of codeword $x_r^i$ can be calculated as $u_r^i = T_{N_r} \cdot x_r^i$. 
The frozen set $\mathcal{F}_r^i$ imposed on the virtual input vector $u_r^i$ depends on the frozen set imposed on the input vector $u$. 
Since $x_r^i$ is a sub-vector of $x$, it is possible to calculate $x_r^i$ directly from $u$ as $x_r^i = u \cdot (T_{N_c}^{(i)} \otimes T_{N_r})$. %, where $A^{(i)}$ is the $i$-th column of matrix $A$. 
%From this we can calculate the relation between input vector $u$ and virtual input vector $u_r^i$ as 
From this, we can calculate the relation between input vector $u$ and virtual input vector $u_r^i$ exploiting the involution property of the transformation matrix of a polar code as 
\begin{align*}
u_r^i & = x_r^i \cdot T_{N_r} \\
 & = u \cdot (T_{N_c}^{(i)} \otimes T_{N_r}) \cdot T_{N_r} \\
 & = u \cdot (T_{N_c}^{(i)} \otimes T_{N_r}) \cdot (I_{N_c} \otimes T_{N_r}) \\
 & = u \cdot (T_{N_c}^{(i)} \cdot I_{N_c}) \otimes (T_{N_r} \cdot T_{N_r}) \\
 & = u \cdot (T_{N_c}^{(i)} \otimes I_{N_r}) 
\end{align*}
For every $l = 0,\dots,N_r-1$, $l$ is in the frozen set $\mathcal{F}_r^i$ only if all the bits of $u$ that are combined to obtain $u_r^i(l)$ are frozen. 
If  matrix $Z$ is defined reshaping $z$ row-by-row, then matrix $Z_c = T_{N_c}^T \ast Z$, where the operator "$\ast$" represents multiplication over $\mathbb{N}$, permits to keep track of the number of unfrozen bits involved in the encoding. 
In practice, each entry of vector $Z_c^{(i,\cdot)}$ represents the number of unfrozen bits used in the calculation of entries of $u_r^i$; only bits having zeros in the corresponding entry of $Z_c^{(i,\cdot)}$ are then frozen. 
%If we call $b_{N_r}^l$ the binary column vector of length $N_r$ having zeros in all entries but in the $l$-th, we have that $l \in \mathcal{F}_r^i$ only if $\mathcal{S}_l \subset \mathcal{F}$, where $\mathcal{S}_l = \arg \max (T_{N_c}^{(i)}  \otimes b_{N_r}^l)$. 
%With slight abuse of notation, we use the $\arg \max$ function to return the set of the indices of a vector for which the entry is one.
%\end{proof}
%\end{proposition}
%
A similar proposition holds for column codes. 
%% calculation can be performed for column codewords $x_C^j$. 
%\begin{proposition}
%\label{prop:frozen_small_c}
%Given the polar code $\mathcal{C}_c^j$ defined on the $j$-th column of $X$, the $l$-th bit belongs to the frozen set $\mathcal{F}_c^j$ only if $\mathcal{S}_l \subset \mathcal{F}$, where 
%\begin{equation}
%\label{eq:frozen_small_c}
%\mathcal{S}_l = \arg \max (b_{N_r}^l \otimes T_{N_c}^{(j)})
%\end{equation} 
%and $b_{N_r}^l$ is the binary column vector of length $N_r$ having one in the $l$-th position and zeros elsewhere.
%\begin{proof}
In this case, $x_c^j = u \cdot (T_{N_c} \otimes T_{N_r}^{(j)})$ and the virtual input vector $u_c^j$ is given by 
\begin{equation*}
u_c^j = u \cdot (I_{N_c} \otimes T_{N_r}^{(j)}) ,
\end{equation*}
and the proof is similar to the one for row codes where $Z_r = Z \ast T_{N_r}$. 
%and the proof is similar to the one for Proposition~\ref{prop:frozen_small_r}. 
%where $Z_r = Z \ast T_{N_r}$, $Z_c = T_{N_c}^T \ast Z$ and the operator "$\ast$" represents multiplication over $\mathbb{N}$. 
%\end{proof}
%\end{proposition}

%\newpage
%\input{Answer}

\end{document}